\documentclass[fleqn,usenatbib]{mnras}
\usepackage{newtxtext,newtxmath}
\usepackage[T1]{fontenc}
\usepackage{ae,aecompl}

\usepackage{graphicx}
\usepackage{amsmath}
\usepackage{amssymb}

\def\dif{\mathrm{d}}
\def\bvec#1{\textrm{\boldmath $#1 $}}

\title[LOFAR limits on DM]{Radio constraints on dark matter annihilation in Canes Venatici I with LOFAR\thanks{Preprint numbers: TUM-1225/19}}
\author[M. Vollmann et al]{Martin Vollmann$^{1}$,\thanks{E-mail: martin.vollmann@tum.de} Volker Heesen$^{2}$, Timothy Shimwell$^{3}$, Martin J. Hardcastle$^{4}$,
\newauthor
Marcus Br\"uggen$^{2}$, G\"unter Sigl$^{5}$ and Huub R\"ottgering$^{6}$
\\
$^{1}$Physik Department T31. James-Franck-Stra\ss e 1, Technische Universit\"at M\"unchen, D-85748 Garching, Germany\\
$^{2}$Hamburger Sternwarte, Gojenbergsweg 112, D-21029 Hamburg, Germany\\
$^{3}$ASTRON, The Netherlands Institute for Radio Astronomy, Postbus 2, 7990 AA Dwingeloo, The Netherlands\\
$^{4}$Centre for Astrophysics Research, School of Physics, Astronomy and Mathematics, University of
Hertfordshire, \\
College Lane, Hatfield AL10 9AB, UK\\
$^{5}$II. Institut f\"ur theoretische Physik, Universit\"at Hamburg, Luruper Chaussee 149, D-22761 Hamburg, Germany\\
$^{6}$Leiden Observatory, Leiden University, PO Box 9513, NL-2300 RA Leiden, The Netherlands
}

\date{Accepted 2020 June 04. Received 2020 May 29; in original form 2019 September 25}
\pubyear{2020}

\usepackage{etoolbox}
\makeatletter
\patchcmd\@combinedblfloats{\box\@outputbox}{\unvbox\@outputbox}{}{%
   \errmessage{\noexpand\@combinedblfloats could not be patched}%
}%
 \makeatother
\begin{document}
\maketitle
\begin{abstract}
Dwarf galaxies are dark matter-dominated and therefore promising targets for the search for
weakly interacting massive particles (WIMPs), which are well-known candidates for dark matter.
Annihilation of WIMPs produce ultra-relativistic cosmic-ray electrons and positrons that emit synchrotron radiation in the presence of magnetic fields. 
For typical magnetic field strengths (few $\mu $G) and $\mathcal
O$(GeV--TeV) WIMP masses, this emission peaks at hundreds of MHz.
Here, we use the non-detection of 150-MHz radio continuum emission from the dwarf spheroidal galaxy Canes Venatici I
with the LOw-Frequency ARray (LOFAR) to derive constraints on the annihilation cross section of WIMPs
into primary electron-positron and other fundamental particle-antiparticle pairs.
Our main underlying assumption is that the transport of the cosmic rays can be
described by the diffusion approximation, thus requiring a non-zero
magnetic field strength with small-scale structure. In particular, by
adopting magnetic field strengths of $\mathcal O(1\,\mu$G) and
diffusion coefficients $\sim 10^{27}~\rm cm^2\,s^{-1}$, we obtain
limits that are comparable with those set by \emph{Fermi} Large Area Telescope using
gamma-ray observations of this particular galaxy. Assuming s-wave
annihilation and WIMPs making up 100 per cent  of the DM density, our benchmark limits exclude several thermal WIMP realisations in the $[2,20]$-GeV mass range. We caution, however, that our limits for the cross section are subject to enormous uncertainties which we also quantitatively assess. In particular, variations on the propagation parameters or on the DM halo can shift our limits up by several orders of magnitude (in the pessimistic scenario).
\end{abstract}

\begin{keywords}
astroparticle physics -- dark matter -- galaxies:dwarf
\end{keywords}

\section{Introduction}
\label{sec:intro}
The $\Lambda$CDM model provides a very successful description of most cosmological
observations \citep[see][for an overview]{2016A&A...594A..13P}. Perhaps most important, cold dark matter can explain the cosmic mass distribution as a result of its gravitational effects. 
Weakly interacting massive particles (WIMPs) are very appealing candidates for dark matter (DM)
and by far the most scrutinised. 
The typical masses of these particles are in the GeV to TeV range and interaction rates that can be accommodated with extensions of the standard model (SM) of particle physics in a rather straightforward manner. 
In the canonical picture, the thermal freeze-out of WIMPs of mass $m_\chi$
occurs at a temperature
$T=T_f\approx m_\chi/20$, which results in a relic mass density relative to the critical density today of \citep{Jungman:1995df}:
\begin{equation}\label{eq:Omega_X}
  \Omega_\chi h^2\sim\frac{3\times10^{-27}~{\rm cm^3\,s^{-1}}}{\langle\sigma_{\chi\bar\chi} \varv\rangle}\,,
\end{equation}
where $\langle\sigma_{\chi\bar\chi} \varv\rangle$ is the total annihilation cross-section multiplied with the relative velocity averaged
over a thermal distribution. Since $\Omega_\chi$, i.e.\ the density parameter of DM in form of WIMPs (henceforth denoted with the greek letter $\chi$) satisfies $\Omega_\chi h^2\la(\Omega_m-\Omega_b)h^2\approx 0.119$ \citep{2016A&A...594A..13P}, equation~(\ref{eq:Omega_X}) puts
a lower limit on the annihilation cross-section at the epoch of decoupling:
\begin{equation}\label{eq:th_X}
  \langle\sigma_{\chi\bar\chi} \varv\rangle\ga\langle\sigma_{\rm th}\varv\rangle\approx 3\times10^{-26}~{\rm cm^3\,s^{-1}},
\end{equation}
where $\langle\sigma_{\rm th}\varv\rangle$ is known as the thermal relic cross-section. The fact that it is of order an
electroweak cross-section is referred to as the `WIMP miracle' \citep{Jungman:1995df}. 

Unfortunately, direct searches of these particles by the means of dedicated direct-detection and collider experiments have yielded only negative results; similarly, no indirect detection of DM by means of astronomical observations has been confirmed. In turn, these experiments and observations have put stringent constraints on several attractive WIMP models \citep{Arcadi:2017kky,Roszkowski:2017nbc}. 

Obviously, the discovery potential of any given DM experiment highly depends on the microscopic properties of the DM model. Diversified detection strategies such as the exploration of the low-frequency radio window for indirect detection of DM are thus essential. Annihilation of WIMP particles 
produces copious amounts of cosmic-ray (CR) electrons and positrons; they emit synchrotron radiation in the presence of magnetic fields. Due to synchrotron and inverse-Compton scattering losses, CR electrons and positrons (CR$e^{\pm}$) are able to propagate only small distances without losing most of its energy \citep[e.g.\ a few hundred parsecs for CR$e^{\pm}$ in the Milky Way;][]{Sigl:2017wfx}. Thus, the otherwise undetectable excess of CR$e^{\pm}$ due to DM annihilation can be probed with radio continuum observations. 

Depending on the DM particle model, this synchrotron emission may be even the strongest signal in the context of multi-messenger astronomy. For example, in scotogenic and leptophilic DM models \citep{Ma:2006km,Fox:2008kb}, or in the context of super-symmetric sneutrino DM models, the DM particles couple to leptons rather than to quarks. These models have such properties that radio continuum observations in the hundreds of mega hertz range stand out as the most promising detection window, as long as the observed targets host strong enough magnetic fields.

Radio continuum observations were applied previously to the DM detection problem. To the best of our knowledge, \cite{2002PhRvD..66b3509T} was the first to make use of radio continuum observations of a dwarf galaxy. They obtained an upper limit for the  $4.9$-GHz flux density of the Draco dSph galaxy from observations with the Very Large Array. Similar recent studies are the ones of  \citet{Regis:2017oet,Leite:2016lsv,Marchegiani:2016xyv,Beck:2015rna,Natarajan:2015hma,Regis:2014tga,Natarajan:2013dsa}. Nevertheless, most of the indirect-detection searches with radio data have focused so far on other types of targets  \citep[mostly the Galactic Centre;][]{Bertone:2001jv,Bertone:2002je,Colafrancesco:2005ji,Bertone:2008xr,Fornengo:2011cn,Fornengo:2011iq,Fornengo:2011xk,Hooper:2012jc,Carlson:2012qc,Storm:2012ty,Cirelli:2016mrc,Storm:2016bfw,Lacroix:2016qpq,McDaniel:2018vam}; the same is true in the context of multi-messenger studies \citep{Regis:2008ij}. 

In this paper, we investigate the ultrafaint dwarf spheroidal (dSph) galaxy Canes Venatici I (henceforth CVnI). It is a satellite galaxy of the Milky Way
at a distance of about 220 kpc from the Sun at (J$2000.0$) R.A.\ $\rm 13^h 28^m 03\fs 5$ and Dec.\ $+33\degr 33\arcmin 21\arcsec$ \citep{zucker_06a}.
It has a mass\footnote{This is the mass that results from integrating the DM density (equation~\ref{eq:density}) within a sphere with a radius of $r_{\rm max}=$ 2.03~kpc, where $r_{\rm max}$ is defined in \citet{2015ApJ...801...74G}.} of $M = 5.6\times 10^8 M_{\sun}$ and an azimuthally-averaged half-light radius of $r_\star= 0.564$~kpc \citep{2015ApJ...801...74G}. 
This dSph galaxy is among the 15 objects considered in the (6~yr) \emph{Fermi} Large Area Telescope (\emph{Fermi}--LAT) search for WIMPs study by \citet{2015PhRvL.115w1301A}.

Our theoretical predictions are based on a standard semi-analytical method that captures the annihilation physics, the diffusive CR propagation and the synchrotron radiation spectrum.
We consider various scenarios for the diffusion coefficient and magnetic field strength.
We do the same with the electron/positron production yields from the annihilation but for brevity only report here the results for exclusive (tree) annihilation into $e^+e^-$ pairs.
The corresponding results for the $\bar b b$, $\tau^+\tau^-$, $W^+W^-$, etc., are included in Appendix~\ref{sec:appannchs}.
This approach has become conventional in the literature as it facilitates the applicability of our results to a wider range of WIMP models. 

Observations with the Low-Frequency Array (LOFAR) are used.
LOFAR is an interferometric radio telescope operating at low
frequencies \citep{van_Haarlem_13a}. We use maps from the preliminary second data release of the LOFAR Two-metre Sky
Survey \citep[LoTSS DR2;][]{Shimwell_17a, Shimwell_19a}, which is a deep 120--168~MHz
imaging survey that will eventually cover the entire northern sky.

This paper is organised as follows. In Section~\ref{sec:pheno}, we discuss the relevant phenomenology for WIMP searches with radio in dwarf galaxies; Section~\ref{sec:obs} presents the LOFAR observations; in Section~\ref{sec:results}, we show our  constraints in the plane defined by the WIMP mass and annihilation cross-section into electron-positron pairs; we then conclude in Section~\ref{sec:concl}. In Appendix~\ref{sec:appannchs}, we include flux predictions and limits due to WIMP annihilation into several combinations of SM particle pairs; in Appendix~\ref{sec:appvariations}, we discuss other sources of uncertainties in our analysis and argue why almost all of them are not significant; Appendix~\ref{a:source_detection} contains details about our source detection tests.

\section{Predictions}
\label{sec:pheno}
In order to obtain our theoretical predictions we follow the approach used in \citet{Leite:2016lsv}, of which we give a brief summary in the following. Microscopic physics is captured by the annihilation cross-section into electrons and positrons, $\langle\sigma \varv\rangle(\chi\chi\to e^\pm{\rm's}+X)$, where the effects of the DM velocity distribution in the observed target are mostly negligible. 
Assuming that the DM is its own antiparticle, the rate at which the electrons and positrons are injected into the dSph galaxy's DM halo is given by:
\begin{equation}\label{eq:injection}
s(\bvec r,E_{e^\pm})=\frac1{2m_\chi^2}\rho^2(\bvec r)\frac{\dif\langle\sigma \varv\rangle}{\dif E_{e^\pm}}\ .
\end{equation}

The DM density $\rho(\bvec r)$ is assumed to be spherically symmetric with respect to the centre of the galaxy and it can be well described by \citep{2015ApJ...801...74G}:
\begin{equation}
\label{eq:density}
\rho(r)=\frac{\rho_s}{\left(\frac{r}{r_s}\right)^{\gamma}\left [1+\left(\frac{r}{r_s}\right )^{\alpha}\right]^{\frac{\beta-\gamma}{\alpha}}} \ ,
\end{equation}
where $\rho_sc_0^2=0.5186~\rm GeV\,cm^{-3}$, $r_s=2.27$~kpc, $\alpha=1.8638$, $\beta=5.9969$, and $\gamma=0.6714$. The variable $r$ is the halo-centric radius and $c_0$ is the vacuum speed of light. 
This set of parameters is consistent with \cite{2015PhRvL.115w1301A}\footnote{In that reference, the parameterisation by \cite{2015MNRAS.451.2524M} is adopted. However, the authors state (see supplemental material) that their results change only by a 10 per cent, had they used the \citep{2015ApJ...801...74G} halo models instead.} to ease comparison.

The quantity $\dif\langle\sigma \varv\rangle/\dif E_{e^{\pm}}$ is
the velocity, angle and spin averaged DM annihilation cross-section
into an electron (or positron) times the relative velocity per unit
energy $E_{e^\pm}$. This quantity depends on the particle physics
model. We can attain some model independence if we  decompose
the cross-sections as the linear superposition of products of {\it hard}
$2\to 2$ cross-sections ($\chi\chi\to e^+e^-$, $\chi\chi\to b\bar b$,
etc.) times the differential $e^{\pm}$ yield from {\it
  final-state} particle cascades. The latter can be obtained from
public Monte--Carlo software packages such as {\tt DarkSUSY}
\citep{Bringmann:2018lay} or {\tt PPPC} \citep{Cirelli:2010xx}. We used {\tt PPPC} but also checked that the results are unaffected had we used  {\tt DarkSUSY}. It follows that:
\begin{equation}
\frac{\dif\langle\sigma v\rangle}{\dif E_{e^{\pm}}}=\sum_{f^+f^-}{\rm BR}(f^+f^-)\langle\sigma \varv\rangle\frac{\dif N_{f^+f^-\to e^\pm+X}}{\dif E_{e^{\pm}}}\ ,
\end{equation}
where $f$ can be any particle of the SM.

Once the electrons and positrons are created by DM annihilation, their propagation can be well described by the diffusion--loss equation:
\begin{equation}\label{eq:diff}
\bvec\nabla\cdot\left[D(\bvec r,E_e)\bvec\nabla n_{e^{\pm}}\right]+b(\bvec r,E_e)\frac{\partial n_{e^{\pm}}}{\partial E}+s(\bvec r,E_e)=0\ ,
\end{equation}
where $D(\bvec r,E_e)$ and $b(\bvec r,E_e)$ are the diffusion and energy-loss coefficients, respectively, and $n_{e^{\pm}}$ is the CR$e^{\pm}$ number density (per unit volume and energy range).

We assume spherical symmetry for simplicity.
While most energy losses are due to the interaction of the electrons and positrons with the ambient electromagnetic field with $b=b_{\rm ICS}+b_{\rm synch}$, where ICS refers to inverse Compton scattering (with photons of the cosmic microwave background), the diffusion is due to the turbulent nature of the magnetic field. The relation between diffusion coefficient and magnetic field structure is complicated and
model dependent. Approximately, in a turbulent field with an rms field strength $B$ and a power spectrum $\delta B^2(k)$ the
diffusion coefficient for an electron is \citep{Sigl:2017wfx}:
\begin{eqnarray}
D(E)&\sim&\frac{E}{3e_0B}\frac{B^2}{\delta B^2(e_0B/E)}\label{eq:diff_coeff}\\
&\sim&3\times10^{22}\left(\frac{E}{{\rm
      GeV}}\right)\,\left(\frac{\mu{\rm G}}{B}\right)\frac{B^2}{\delta B^2(e_0B/E)}\,\frac{{\rm cm}^2}{{\rm s}}\,,\nonumber
\end{eqnarray}
where $e_0$ is the elementary charge; note that $r_L=E/(e_0B)$ is the Larmor radius of an electron in a magnetic field $B$. The first factor in equation~(\ref{eq:diff_coeff}) is referred to as the Bohm limit for diffusion. Because of the second factor, the real diffusion
coefficient is always larger than the one obtained for the Bohm limit. 

In the Milky Way, CR abundance measurements, such as of the boron-to-carbon ratio, give diffusion
coefficients $D_0\equiv D(1~{\rm GeV})$ of order  $10^{28}~\rm cm^2\,s^{-1}$ \citep[e.g.][]{Korsmeier:2016kha}, with similar values found in external galaxies for the CR $e^{-}$ from spectral ageing \citep[e.g.][]{heesen_19a}. These values are
a factor of $\sim\,10^7$ larger than for the Bohm limit. This is probably due to the small fractional magnetic field power $\delta B^2(10^{-6}\,{\rm pc})/B^2\sim10^{-7}$ at the Larmor radius $r_L\lesssim10^{-6}\,$pc of a GeV-electron.
If the magnetic power-spectrum is a power-law with a slope of $n\sim-1$ at scales below the field coherence length $l_c$, one has
$\delta B^2(k)\sim(kl_c)^n$. For $l_c\sim10$~pc, this results in the right order of magnitude. Given the large uncertainties of the magnetic field structure, we treat the diffusion coefficient and magnetic field strength as independent
parameters for our purposes, as long as $D_0\sim10^{27}~\rm cm^2\,s^{-1}$ and $B\sim 1\,\umu$G hold within one to two orders of magnitude, which we assume throughout this work. 

For CR$e^{-}$ in the GeV-regime, the diffusion coefficients in late-type spiral galaxies are of the order of $10^{28}\,\rm cm^2\,s^{-1}$ \citep{murphy_08a,heesen_19a}; this can be derived from their kpc-transport lengths and radiation lifetimes of a few 10~Myr. Galaxies seen edge-on can have large radio haloes, so that their diffusion coefficients may be even larger with values of a few $10^{29}~\rm cm^2\,s^{-1}$. However, the most likely explanation is that they possess galactic winds and the CRe transport is advection dominated \citep{heesen_18b}. In dwarf irregular galaxies, the diffusion coefficients are smaller with values of the order of $10^{27}$ \citep{murphy_12a} or possibly even as low as $10^{26}~\rm cm^2\,s^{-1}$ \citep{heesen_18c}. However, these dwarf galaxies have star formation, so that their magnetic fields and diffusion coefficients are probably not relevant in the case of dSph galaxies such as CVnI.
Nevertheless, we choose $D_0=10^{27}~\rm cm^2\,s^{-1}$ as our benchmark value for CR$e^{\pm}$ with an energy of 1~GeV.

Concerning the magnetic field strength, the situation is much more uncertain though. 
We require the magnetic energy density to be in equipartition with the CR energy density within 2 orders of magnitude. For our radio continuum sensitivity, we expect an equipartition magnetic field strength of $\approx 1~\mu \rm G$ \citep[for a $e^{\pm}$ plasma,][]{beck_05a}. Should the magnetic field strength be much smaller than that, the CRs cannot be reasonably confined in the galaxy and would leave with a speed comparable to the speed of light. To prevent our assumption of diffusion to break down, a lower (pessimistic) magnetic field strength of $0.1$~$\mu\rm G$ is reasonable.

In the following, we use this parametrisation of the energy-dependency of the diffusion coefficient:
\begin{equation}\label{eq:diff_para}
D(E)=D_0\left(\frac{E}{{\rm GeV}}\right)^\delta\,,
\end{equation}
with $\delta=1+n$ in the model above. We will take $\delta=1/3$ which is supported by observations in the Milky Way \citep{Korsmeier:2016kha}.
As a first approximation, we further assume that the value of the diffusion coefficient approaches infinity at radius $r_h$ [$D(E) \rightarrow \infty$], and is homogeneous inside the sphere of the same radius -- an assumption that has become standard in the literature \citep{Colafrancesco:2005ji,Colafrancesco:2006he,McDaniel:2017ppt}. Then, by adopting a semi-analytical approach, it is possible to solve the diffusion--loss equation in terms of Green's functions \citep{Vollmann:2020undprp}.

\begin{figure}
\center{\includegraphics[width=0.42\textwidth]{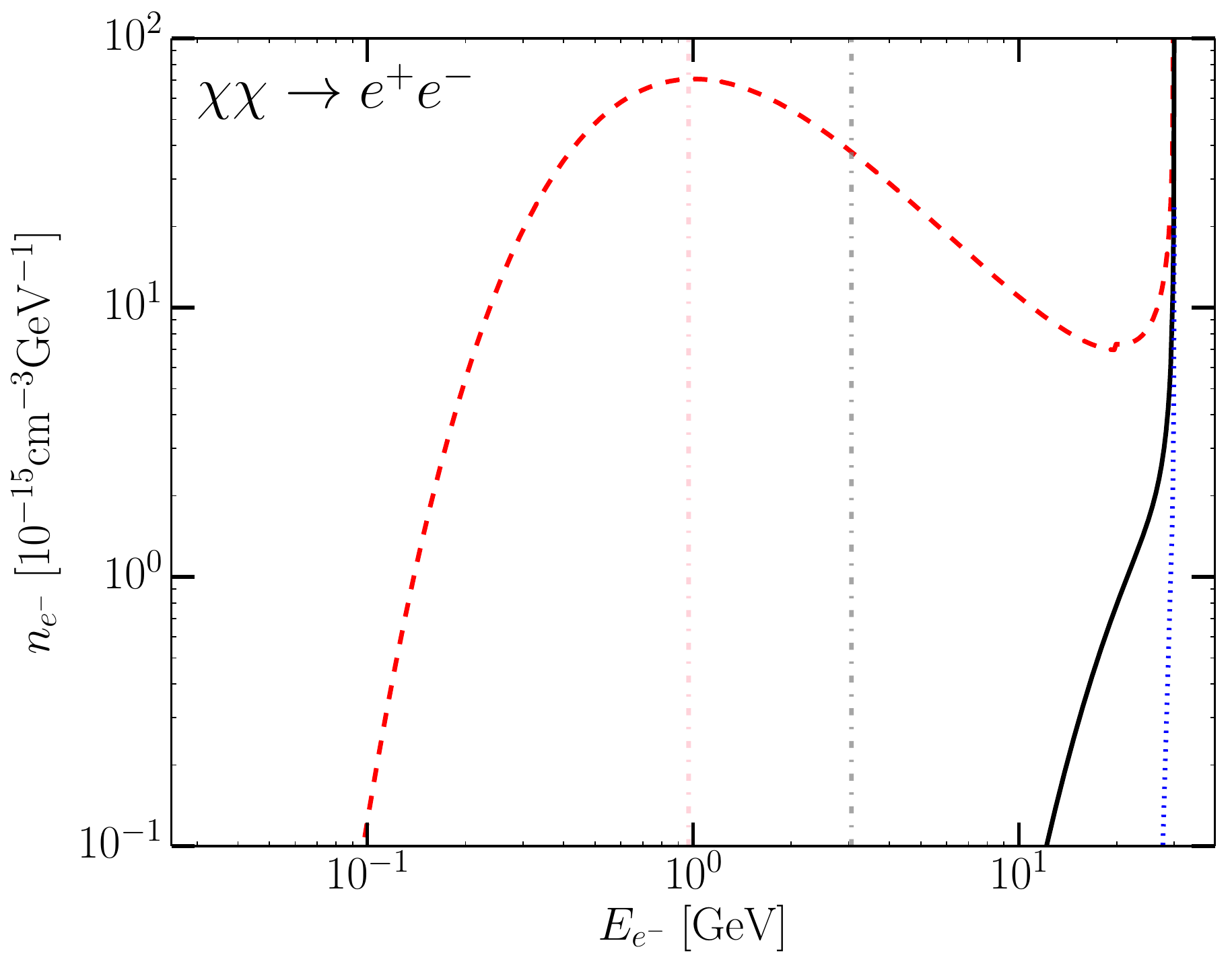}}
\caption[...]{Electron/positron-density spectrum at the centre of the dSph galaxy resulting from the annihilation of 30-GeV  DM particles. We assume $B=10~\umu$G and $D_0=10^{26}~\rm cm^2\,s^{-1}$ (red-dashed line); $B=1~\umu$G and $D_0=10^{27}~\rm cm^2\,s^{-1}$ (black-solid line); and $B=1~\umu$G and $D_0=10^{28}~\rm cm^2\,s^{-1}$ (blue-dotted line). Vertical lines mark the electron energy $E_c$ at which $\nu_c=150$~MHz for the two magnetic field strengths considered. See text for details.}
\label{fig:edensity}
\end{figure}

The resulting CR$e^\pm$ distribution that is originates from DM annihilation is shown in Fig.~\ref{fig:edensity}. It is evident that the distribution carries rather distinctive features. It becomes infinite at the mass of the DM particles, which is a consequence of the monochromatic energy distribution of the emitted electron--positron pairs per annihilation. It also features a low-energy cut-off at some specific $e^\pm$ energy, which strongly depends on the diffusion coefficient and the magnetic field strength. Electrons and positrons at lower energies than that have diffused away from the dwarf galaxy. 

Since the CR$e^{\pm}$ injection-rate density (equation~\ref{eq:injection}) peaks at the centre of the dwarf galaxy and falls off towards the edges, we
use $n_{e^\pm}(r_h)=0$ at $r_h=$~1~kpc as boundary condition, adopting a radius of $2r_\star$. 
We verified that the (computationally favourable) boundary condition $n_{e^\pm}(r_h)=0$ is compatible with the physical one: $D(r_h,E_e)(\partial n_{e^\pm}/\partial r)(r_h)=c_0n_{e^\pm}(r_h)$ is fulfilled in all cases considered. 
 
The radio emissivity associated with this synchrotron radiation is:
\begin{equation}\label{eq:emissivity}
j_\nu(\bvec r)=\int\dif E_{e^-}2\,n_{e^-}({\bf r},E_e)P_\nu(E_{e^-},B)\ ,
\end{equation}
where $P_\nu(E_{e^-},B)$ is the pitch-angle averaged emitted power of a single electron in the presence of a magnetic field with rms strength $B$.
The factor of 2 accounts for the fact that 
for CP-invariant models for DM as many positrons as electrons are produced in every annihilation. Then $P_\nu$ can be conveniently  written as \citep{Leite:2016lsv}:
\begin{equation}
P_\nu(E_e)=\frac{9\sqrt3}{8\pi}\frac{b_{\rm synch}(E_e,B)}{\nu_c(E_e,B)}F\left(\frac{\nu}{\nu_c(E_e,B)}\right)\ ,
\end{equation}
where $b_{\rm synch}$ is the synchrotron energy-loss rate, $F(x)$ is defined as \citep{1988ApJ...334L...5G}:
\begin{equation}
F(x)=6x^2\left[K_{4/3}(x)K_{1/3}(x)-\frac35\left(K_{4/3}^2(x)-K^2_{1/3}(x)\right)\right]\ ,
\end{equation}
and $\nu_c=3e_0BE^2/(4\upi m_e^2)$ is the critical frequency of the synchrotron radiation spectrum. In Fig.~\ref{fig:edensity}, we indicate the characteristic energy $E_c$ that results from inverting this equation and plugging in the observation frequency of $\nu_c=$~150~MHz. Electrons with energies smaller than this do not significantly emit synchrotron radiation at the observation frequency. The predicted radio continuum intensity is then:
\begin{equation}
I_\nu=\int_{\rm LoS}j_\nu[\bvec r(l)] \dif l\,,
\end{equation}
which is the line-of-sight (LoS) integral of equation~(\ref{eq:emissivity}).

\begin{figure}
\center{\includegraphics[width=0.42\textwidth]{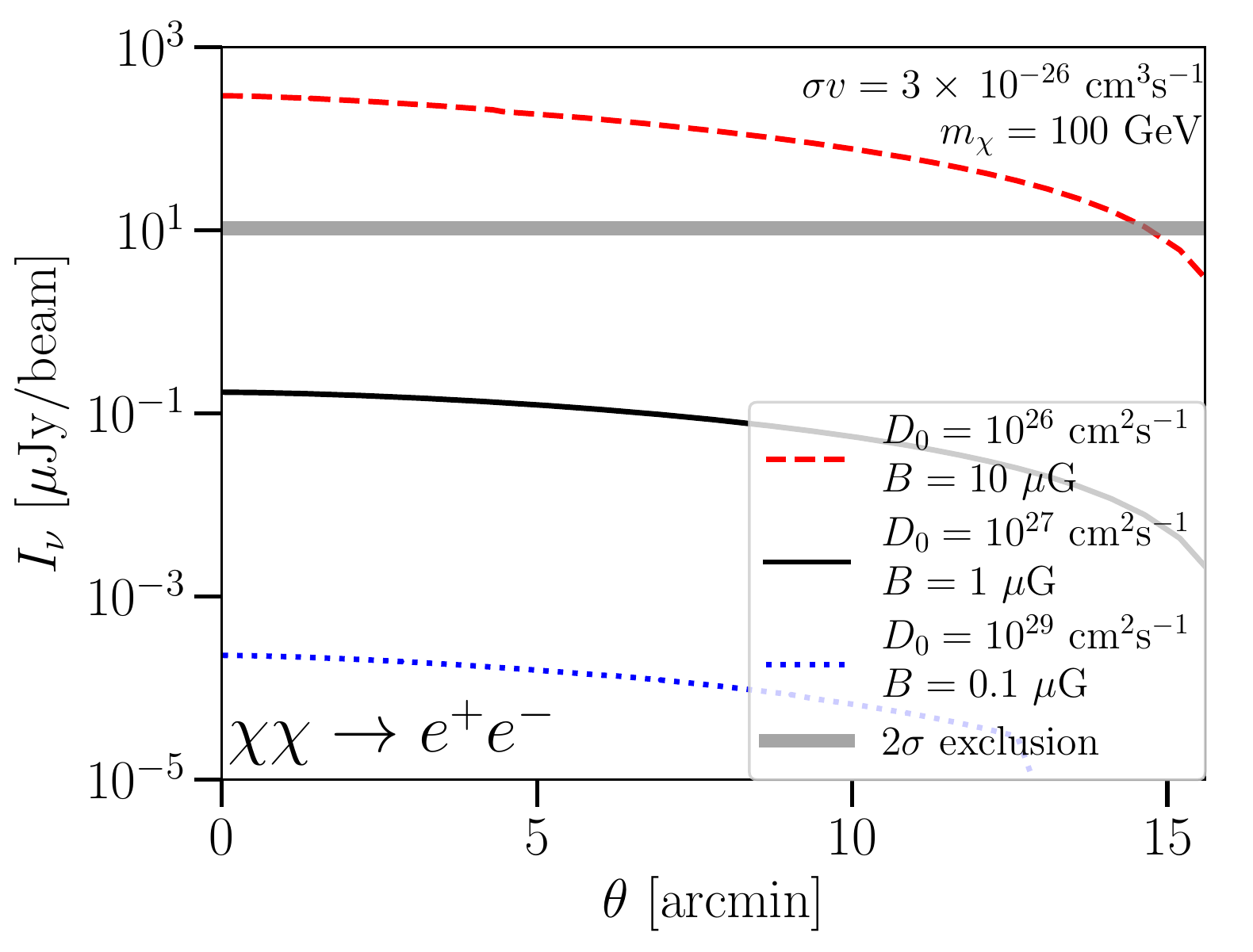}}
\caption[...]{Predicted radial profile of the 150-MHz radio continuum intensity for various combinations of
diffusion coefficients $D_0$ and rms magnetic field strengths $B$. We assume a specific WIMP mass and annihilation cross-section into $e^+e^-$
as indicated. The grey line indicates the maximum flux in the {\it spherical tophat} model discussed in Section~\ref{sec:results} that is excluded at level of $2\sigma=11~\mu \rm Jy\,beam^{-1}$ with our observations.}
\label{fig:flux}
\end{figure}

Figure~\ref{fig:flux} shows the predicted 150-MHz radio continuum radial intensity profile as a function of projected radius expressed by the apparent angle $\theta$; we assume 100-GeV DM particles that annihilate into electron--positron pairs with the `thermal' cross section.  Three different scenarios are considered: (1) the \emph{optimistic} scenario, where the magnetic field is strong (10~$\mu$G) and highly turbulent (and hence the diffusion coefficient is small $D=10^{26}\,\rm cm^2\,s^{-1}$ for $E_{e^\pm}=1$~GeV); (2) our \emph{benchmark} scenario ($B=1\,\umu$G, $D_0=10^{27}~\rm cm^2\,s^{-1}$); and (3) the \emph{pessimistic} scenario, where energy losses can be neglected in equation~\eqref{eq:diff} ($B=0.1~\umu$G and $D_0=10^{29}~\rm cm^2\,s^{-1}$). In scenario (3), neglecting energy losses is justified since the diffusion length within the CR$e^{\pm}$ lifetime exceeds the system size. While the predicted intensities are different for our three scenarios, their spatial distributions have similar shapes with a full width at half-maximum (FWHM) of $\approx$8~arcmin (for $r_h=1$~kpc). This is the area from which the radio continuum signal is expected to be seen. The maximum intensities and FWHMs will be of course affected by the choice of $r_h$.
See Appendix \ref{sec:appvariations} for a quantitative account of these parameter choice variations.

Notice that the optimistic case assumes $B=10\,\mu$G, which can rightfully be considered as extreme for a presumably quiet, non-starforming dSph galaxy such as CVnI. We include it for completeness because it resembles most closely the limit in which the time-scale of diffusion is much larger than the corresponding CR$e^{\pm}$ lifetime; the result serves also as a verification of the correctness of our predictions. However, even such a large value is still consistent with measurements of Faraday rotation.  \citet{2015A&A...575A.118O} estimates the extragalactic contribution to the rotation measure (RM) at the position of CVnI to be RM~$\approx1\pm6~\rm rad\,m^{-2}$; this value is compatible with $B=10\,\mu$G if typical gas densities $\lesssim 10^{-3}$~cm$^{-3}$ are assumed. Indeed, \citet{spekkens_14a} provide an upper limit for the H\,{\sc i} mass of $1200~\rm M_{\sun}$, which translates into an even lower neutral hydrogen density of $10^{-5}~\rm cm^{-3}$, corroborating the assumption of low gas densities.

 Moreover, if magnetic fields were amplified in the past as a result of star formation, the field strength could be as high as a few 10~$\mu$G if we take nearby dwarf irregular galaxies as the equivalent of such a galaxy \citep{hindson_18a}. The magnetic diffusivity is very small in the interstellar medium, so that remnant magnetic fields could be preserved into today's Universe. During the fragmentation of the gaseous disc, the gas slips along the magnetic field lines as a result of the Parker instability to form molecular clouds \citep{koertgen_19a}. The magnetic field is left behind. 

\section{LOFAR Observations}
\label{sec:obs}
\begin{figure}
\center{\includegraphics[width=0.5\textwidth]{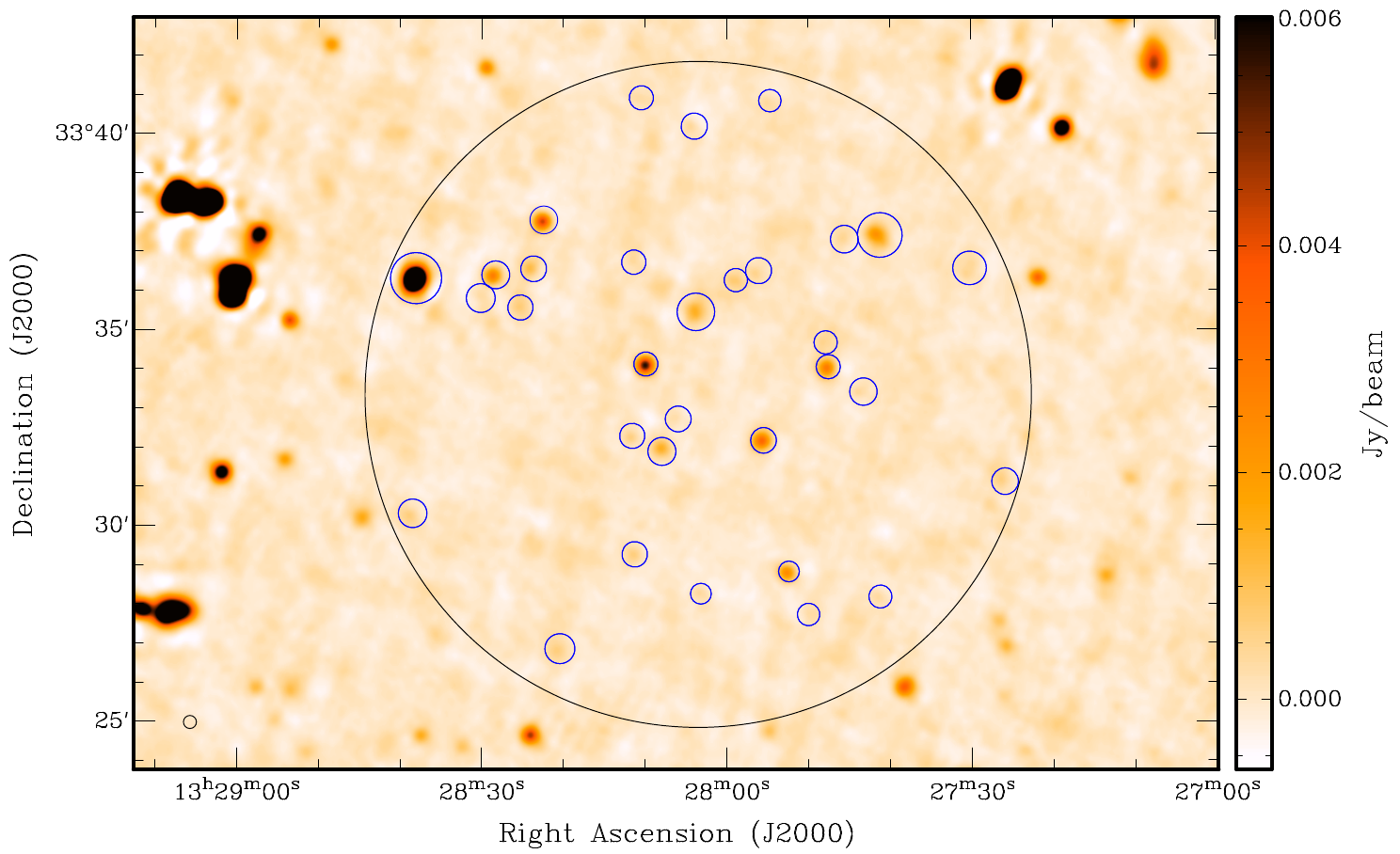}}
\caption[...]{LoTSS 150-MHz map of the region around CVnI at 20~arcsec
  FWHM angular resolution. We show the intensity at linear stretch between $-0.6$ and $6~\rm mJy\,beam^{-1}$. The large black circle indicates the area in
  which we integrated the intensity to measure the flux density, and the 32 small blue circles
  indicate the sources that we subtracted. The small black circle in the bottom-left corner shows the synthesized beam.}
\label{fig:map}
\end{figure}

\begin{figure}
\center{\includegraphics[width=0.5\textwidth]{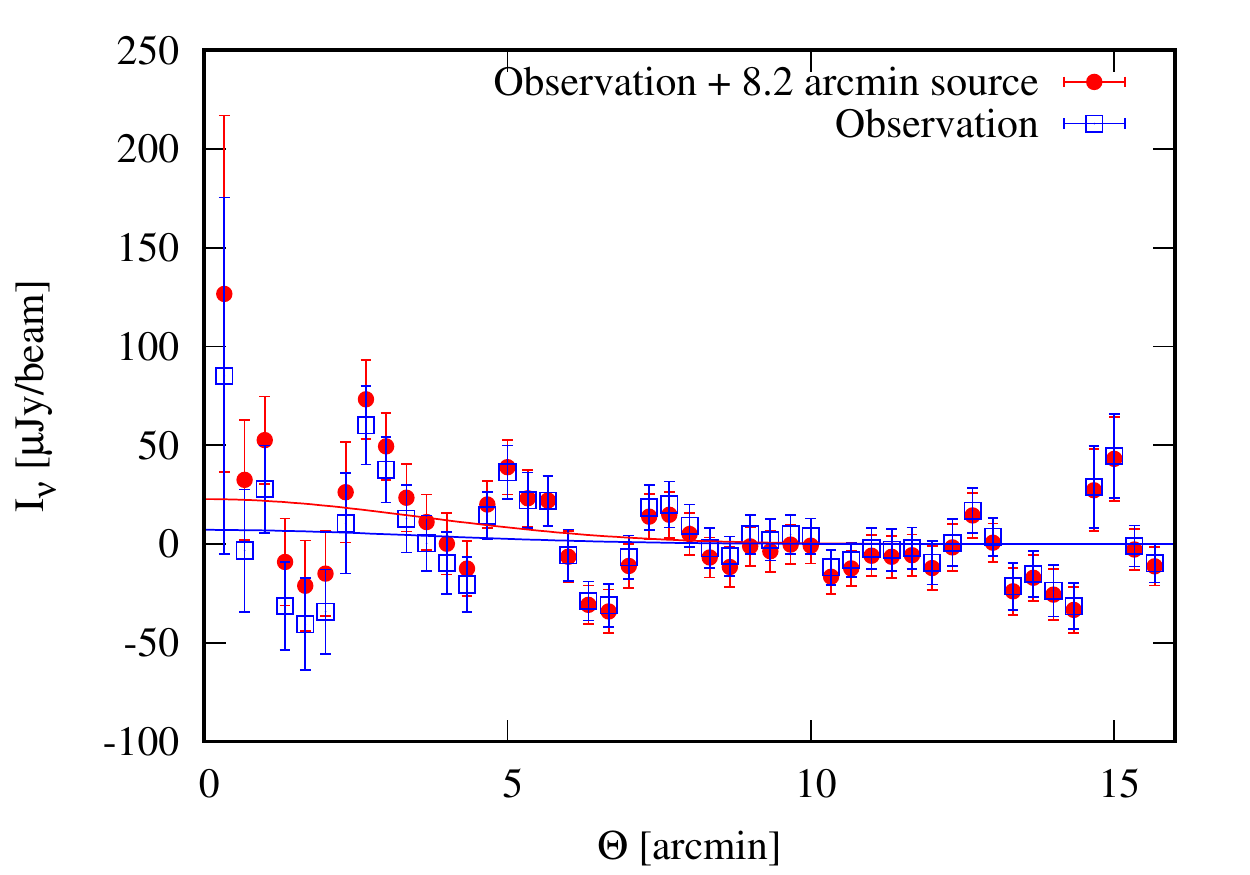}}
\caption[...]{Radial profile of the 150-MHz intensity in the region around CVnI at 20~arcsec
  FWHM angular resolution. Point-like sources have been subtracted. Open blue data points show the measured profile. Filled red data points show the measured profile with a fake 20-mJy source with $\rm FWHM=8.2~arcmin$ (equivalent to an amplitude of $a=32~\rm \mu Jy\,beam^{-1}$) added that resembles the expected DM annihilation profile. Solid lines show the best-fitting Gaussian intensity profiles. Whereas the measured data shows no excess flux (blue line), the fake source is well reproduced by the fit (red line).}
\label{fig:profile}
\end{figure}

\begin{figure}
\center{\includegraphics[width=0.5\textwidth]{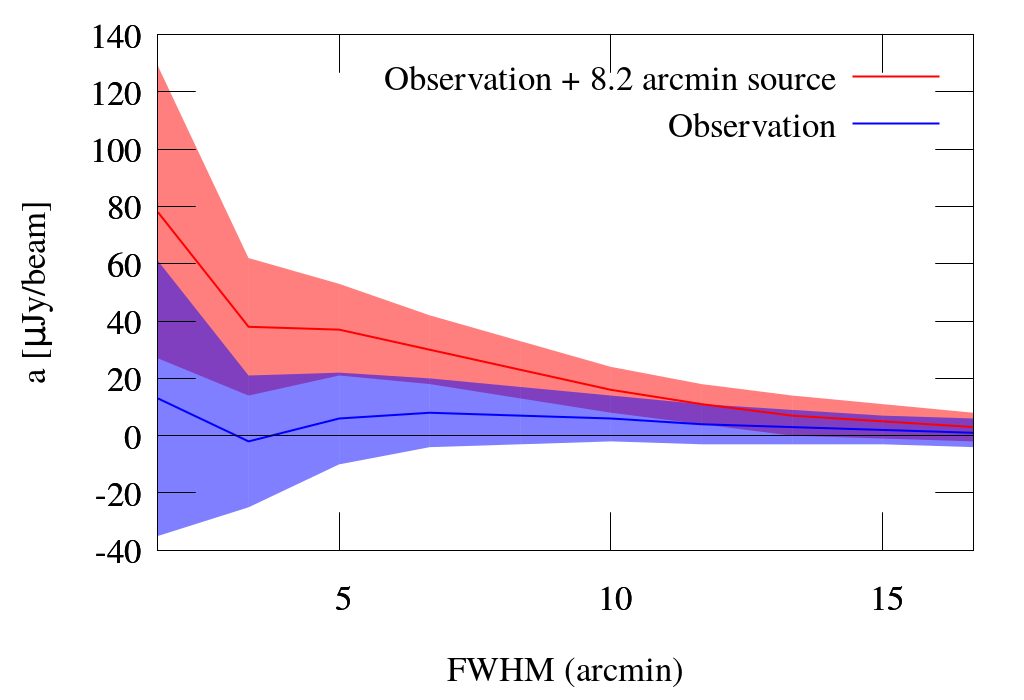}}
\caption[...]{Best-fitting amplitude $a$ for a Gaussian function fitted to the radial intensity profile as function of the assumed value for $b$ (here expressed by FWHM). Lines show the best-fitting amplitudes for the data with a fake 20-mJy source with $\rm FWHM=8.2~arcmin$ inserted (red) and the control data with no source inserted (blue). Shaded areas indicate 1$\sigma$ uncertainties with 1 degree of freedom.}
\label{fig:sigma_500}
\end{figure}

We use LoTSS 150-MHz data in order to search for the possible radio continuum emission from CVnI. First, we use a preliminary LoTSS DR2 150-MHz map at an angular resolution of 6~arcsec in order to identify point-like background sources. With {\small PYBDSF} \citep[Python Blob Detection and Source Finder;][]{mohan_15a} we identified these sources and subtracted them from the $(u,v)$ data. Then we re-imaged and deconvolved the $(u,v)$ data with {\small WSCLEAN} v$2.7$ \citep{offringa_14a} at 20~arcsec FWHM angular resolution. In order to reduce foreground contamination from the Milky Way, we also applied a lower $(u,v)$-cut of 160~$\lambda$, so that we are sensitive to emission on angular scales of up to $\approx$21~arcmin; this is well above the size of the galaxy, which has an
optical radius of $8.5~\rm arcmin$ corresponding to 540~pc.

Figure~\ref{fig:map} shows the radio continuum intensity at 150~MHz in area of approximately $26\times 20$~arcmin$^2$ centered on CVnI, prior to the subtraction of sources. A number of unresolved point-like
sources can be seen, 32 of which are located within the $8.5$-arcmin
radius. We detect no diffuse emission within this radius, which would be
the expected morphology for DM-generated radio continuum emission; hence, we assume point-like sources to be unrelated
to the galaxy. 
The correct way
to ascertain the non-existence of diffuse emission is to integrate the
intensity and check whether it is consistent with zero. For this we need
to take rms map noise into account, which is $\sigma_{150~\rm MHz}=130~\rm \umu Jy\,beam^{-1}$. Within a radius
of $8.5$~arcmin the flux density at 150~MHz is $S_{150~\mathrm{MHz}}=(-3.6\pm5.5)~$mJy after subtraction of
the point-like sources, which contribute $89.5$~mJy in total. The resulting residual flux density is consistent with zero. This equates to a 2$\sigma$ uncertainty for the intensity of only $6~\mu\rm Jy\,beam^{-1}$. However, this estimate is too optimistic since it neglects the effects of the deconvolution. In particular, such large extended sources may not be picked up by {\small WSCLEAN} algorithm and hence may not be deconvolved.

In order to test the estimated uncertainty, we inserted fake Gaussian sources into the $(u,v)$ data and then imaged and deconvolved the data in the same way as the original data. The intensity model in Fig.~2 has a FWHM of $8.2$~arcmin; hence, we inserted Gaussian sources with such a FWHM into the data. The model intensity distribution depends on the radial size of the DM halo, which is not known, so we varied the FWHM between $4.1$ and $16.4$~arcmin ($r_h=0.5~\rm kpc$ and $2$~kpc, respectively). Then we investigated the radial intensity profile as shown in Fig.~\ref{fig:profile}. As one can see, the intensities in with the source added are only slightly higher than the observations without the fake source. What dominates are the local fluctuations in the map that decrease at large angular radii, where the integration area increases. A noticeable side-effect from inserting fake sources into the data is that since such large sources are not picked up by the {\small CLEAN} algorithm; the data outside of 4~arcmin radius are lower with the fake source inserted. The reason for this is that the source is not deconvolved and hence the sidelobe will result in a negative intensity contribution.

We found that the fake sources could be detected at 1--2$\sigma$ significance if they are co-spatial with the optical centre of the galaxy. An example for this testing is shown in Fig.~\ref{fig:sigma_500}, where we have inserted a Gaussian fake source with of 20-mJy flux density and $\rm FWHM=8.2~\rm arcmin$. Hence, this source can be described by:
\begin{equation}
    I_{\nu} = a \times \exp\left (\frac{-\Theta^2}{2b^2} \right ),
\end{equation}
where $a=32~\rm \mu Jy\,beam^{-1}$ and $b=3.5~\rm arcmin$ $\lbrace b=\rm FWHM/[2\sqrt{2\ln(2)}]\rbrace$. We now have fitted this radial intensity profile with Gaussian searching for the best-fitting amplitude $a$ with various values of $b$ (and hence FWHM) fixed. In Fig.~\ref{fig:sigma_500}, the recovered amplitude with 1$\sigma$ uncertainty is presented. We recover approximately the correct amplitude when setting $\rm FWHM=8.2~arcmin$ with $a=23\pm 10~\rm \mu Jy\,beam^{-1}$ (red line in Fig.~\ref{fig:profile}). But we detect the source with $1\sigma$ significance up to a radius of 13~arcmin. In contrast, the data without sources does not show anywhere a 1$\sigma$ significant detection. So this search in the observed radial intensity profile for Gaussian distributions confirms the non-detection. For the other inserted sources see Appendix~\ref{a:source_detection}. We conclude that between $r_h = 0.5~\rm kpc$ and $r_h=2~\rm kpc$ we can rule out a signal from DM annihilation with a peak intensity of $32~\rm \mu Jy\,beam^{-1}$ at $2\sigma$ significance. Averaged across the size of the galaxy this equates to an average intensity of 11~$\mu\rm Jy\,beam^{-1}$.

\section{WIMP Constraints}
\label{sec:results}
Bearing in mind that a thorough search for the DM signal discussed in Section \ref{sec:pheno} demands the employment of advanced data-analysis methods, we will content ourselves by \emph{estimating} limits on $\langle\sigma \varv\rangle$ for several annihilation channels.

\begin{figure}
\center{\includegraphics[width=0.42\textwidth]{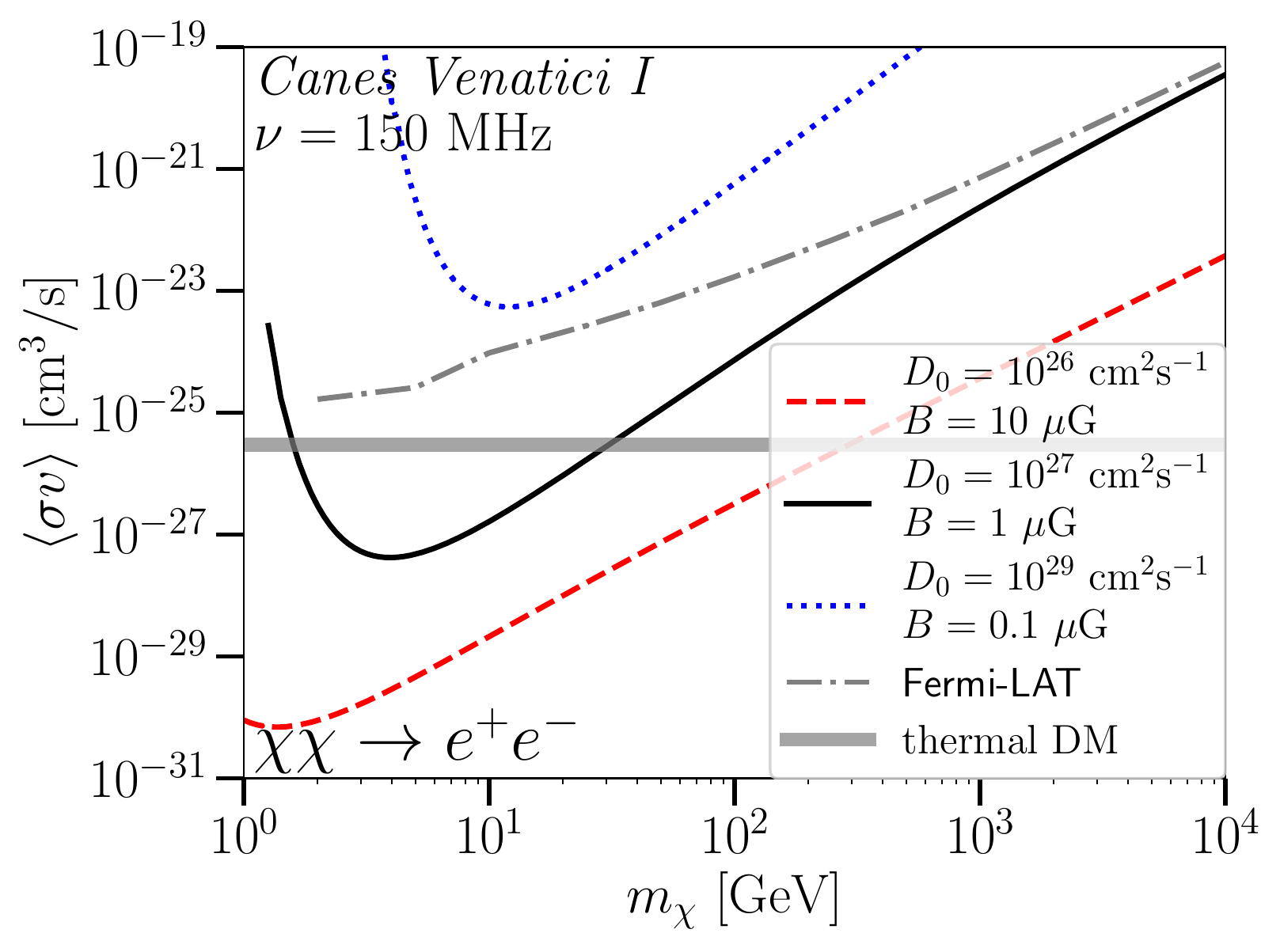}}
\caption[...]{Resulting constraints on the WIMP annihilation cross-section into $e^+e^-$ as a function of WIMP mass $m_\chi$,
for various combinations of diffusion coefficients and magnetic field strengths. All values are for the specific case of CVnI only. Particularly, \citet{2015PhRvL.115w1301A}, using \emph{Fermi}--LAT data, have derived lower limits for the combined constraints from their galaxy sample.
}
\label{fig:eelimits}
\end{figure}

Concretely, we use the 2$\sigma$ limit on the maximum flux of a signal with a spherical tophat shape that can be extracted from a noisy image. This can be obtained analytically and is given by \citep{Leite:2016lsv,Vollmann:2020undprp}:
\begin{equation}\label{eq:obs_noise}
I_{150~\rm MHz}^{\textrm{excluded at }2\sigma}=1.64\frac{\sigma_{150~\rm MHz}}{\sqrt{N_{\rm beams}}} \ ,
\end{equation}
where $N_{\rm beams}$ is the effective number of beams that are required to image the tophat signal \citep{Vollmann:2020undprp}. 
In practice, we estimated $N_{\rm beams}$ as the ratio between the solid angle of a cone whose major and minor axes follow the DM distribution of CVnI and such that it contributes half of the total flux, to the solid angle corresponding to the Gaussian synthesized beam of the LOFAR map. In principle, $N_{\rm beams}$ depends on the annihilation channel, DM particle mass, and diffusion coefficient. However, this dependence is weak and we find that the FWHM amounts 
to 8.2~arcmin, corresponding to $N_{\rm beams}=710$. 
Figure~\ref{fig:eelimits} shows the resulting constraints on the WIMP annihilation cross-section into $e^+e^-$ as a function of mass
resulting from comparing our radio continuum intensity predictions with the observational upper limit.

As a cautionary remark, the cross-section constraints obtained here are on \emph{present-time} DM annihilation and they should thus not directly be compared with the thermal freeze-out annihilation
cross-section $\langle\sigma_{\rm th}\varv\rangle$ that is relevant for the relic density. WIMP annihilation probed today is sensitive to relative velocities $\varv/c_0\sim10^{-3}$. In contrast,
thermal freeze-out is governed by the total annihilation cross-section at velocities $\varv/c_0\sim0.3$.
Therefore, a numerically much stronger constraint than $\langle\sigma_{\rm th}\varv\rangle$ can be reconciled with thermal freeze-out by either assuming
an annihilation cross-section that decreases with $\varv$, such as in higher partial waves, or by assuming a branching ratio into electrons and positrons that is
much smaller than unity; a combination of both is of course also possible.

\section{Conclusions}
\label{sec:concl}
Radio continuum observations of nearby dwarf galaxies offer the possibility to indirectly detect emission from dark matter, such as expected from the annihilation of WIMPs. For our educated guesses for the magnetic field strengths and the mass ballpark of the WIMP (few GeV to a few TeVs), the synchrotron emission is expected to peak in the low-frequency radio continuum regime. 

In this paper, we have used a 150-MHz map from the LOFAR Two-metre Sky Survey (LoTSS) to search for radio continuum emission from Canes Venatici I, a dSph satellite galaxy of the Milky Way. We do not detect any diffuse emission, allowing us to put constraints on the DM annihilation cross-sections into secondary electron and positron cascades for the generic DM models; we pay particular attention to primary hard electron--positron pairs from the $2\to 2$ annihilation process. For WIMP masses of
$2~{\rm GeV}\lesssim m_\chi\lesssim 20~{\rm GeV}$, the upper bounds on the primary $e^+e^-$ process from our benchmark scenario (2)
are smaller than the total thermal relic cross-section. In the [2~GeV,~1~TeV] energy interval, our benchmark limits are more stringent than the ones by \emph{Fermi}--LAT observations of CVnI (see Fig.~\ref{fig:eelimits}). A similar situation occurs when the electron and positrons from DM annihilation are produced by particle cascades from other leading scenarios of hard processes, such as $\chi\chi\to\tau^+\tau^-$, if stronger
assumptions on the magnetic field strength and diffusion coefficients are made.  
 
 This proof-of-concept study is the first of its kind at the low frequencies probed by LOFAR. Since the predicted CR electron/positron distribution in Fig.~\ref{fig:edensity} is fairly different from astrophysical spectra, the associated synchrotron signal benefits from distinctive features that
can be explored in more ambitious multi-frequency and multi-object studies. The main limitation for this kind of work is the unknown magnetic field structure in dwarf galaxies with little or no star formation. The cosmic ray diffusion approximation requires magnetic fields with small scale turbulence-like structure and the magnetic energy density should be order of magnitude comparable to the cosmic-ray energy density. Even if the diffusion approximation holds, our limits for the WIMP annihilation cross-section are two orders of magnitude higher if we assume more pessimistic parameters for the magnetic field strength and diffusion coefficient.

\section*{Acknowledgments}
We thank the anonymous referee for their insightful report and suggestions that improved the manuscript. MV and GS would like to thank the Max Planck Institute for Physics and Georg Raffelt in particular for their hospitality and feedback. While completing this work, we became aware of two papers \citep{Kar:2019hnj,Kar:2019cqo}, where similar studies are performed with different telescopes and targets. Their results are similar to those presented here.

This work is partly funded by the Deutsche Forschungsgemeinschaft under Germany's Excellence Strategy EXC 2121 `Quantum Universe' 390833306 and the Collaborative Research Center `Neutrinos
and Dark Matter in Astro- and Particle Physics' (SFB 1258). MJH acknowledges support from the UK Science and Technology Facilities Council (grant ST/R000905/1). This research has made use of the University
of Hertfordshire high-performance computing facility
(\url{https://uhhpc.herts.ac.uk/}) and the LOFAR-UK compute facility,
located at the University of Hertfordshire and supported by STFC (ST/P000096/1).

LOFAR, the Low Frequency Array, designed and constructed by ASTRON, has
facilities in several countries, which are owned by various parties
(each with their own funding sources), and are collectively operated
by the International LOFAR Telescope (ILT) foundation under a joint
scientific policy. The ILT resources have benefited from the
following recent major funding sources: CNRS-INSU, Observatoire de
Paris and Universit\'e d'Orl\'eans, France; BMBF, MIWF-NRW, MPG, Germany;
Science Foundation Ireland (SFI), Department of Business, Enterprise
and Innovation (DBEI), Ireland; NWO, The Netherlands; the Science and
Technology Facilities Council, UK; Ministry of Science and Higher
Education, Poland.

The data underlying this article are available on the website of the Centre de Donn{\'e}es 
astronomiques de Strasbourg (CDS; \href{http://cds.u-strasbg.fr}{http://cds.u-strasbg.fr}). 

\appendix

\section{Further annihilation channels}
\label{sec:appannchs}
\begin{figure*}
\includegraphics[width=.49\linewidth]{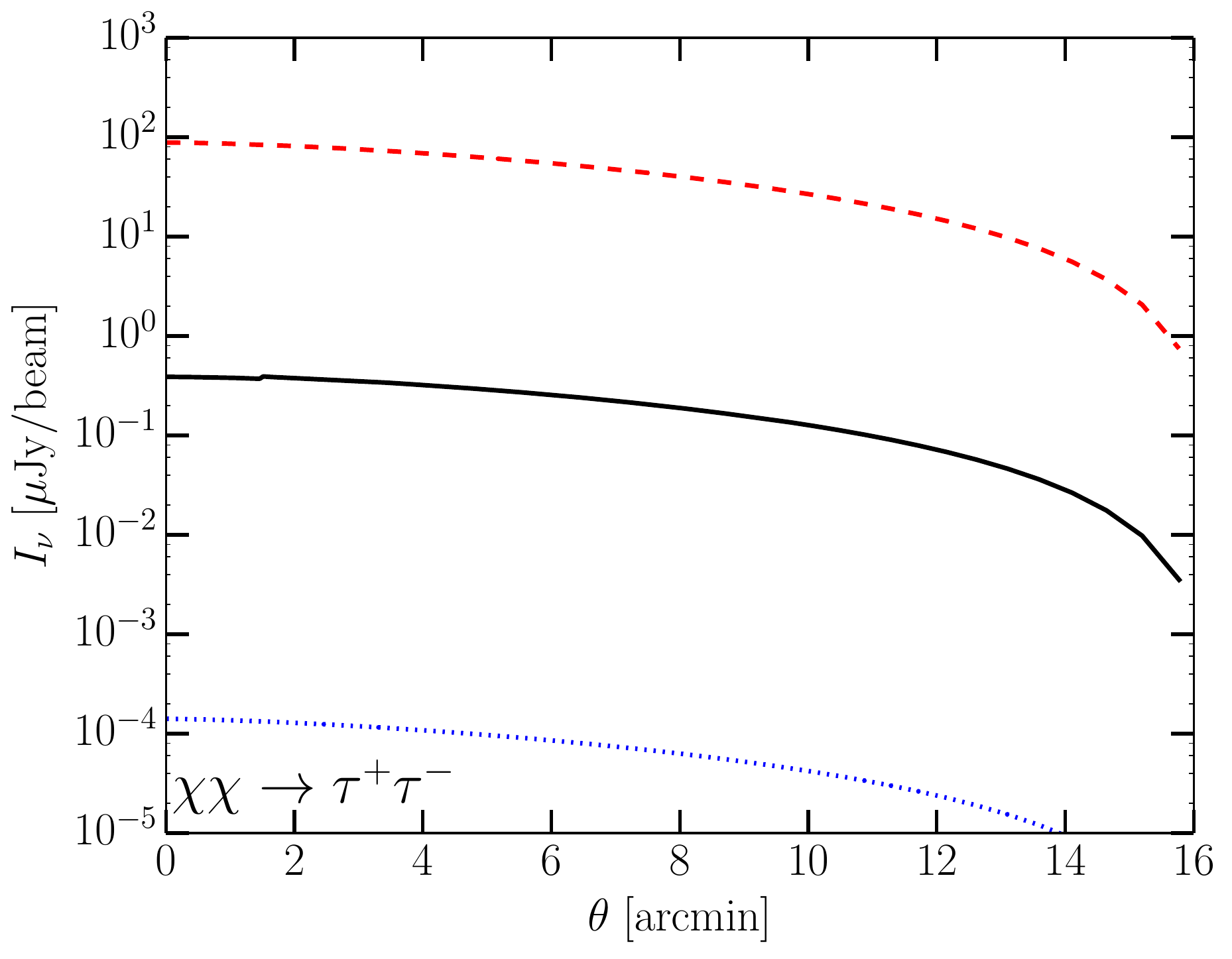}
\includegraphics[width=.49\linewidth]{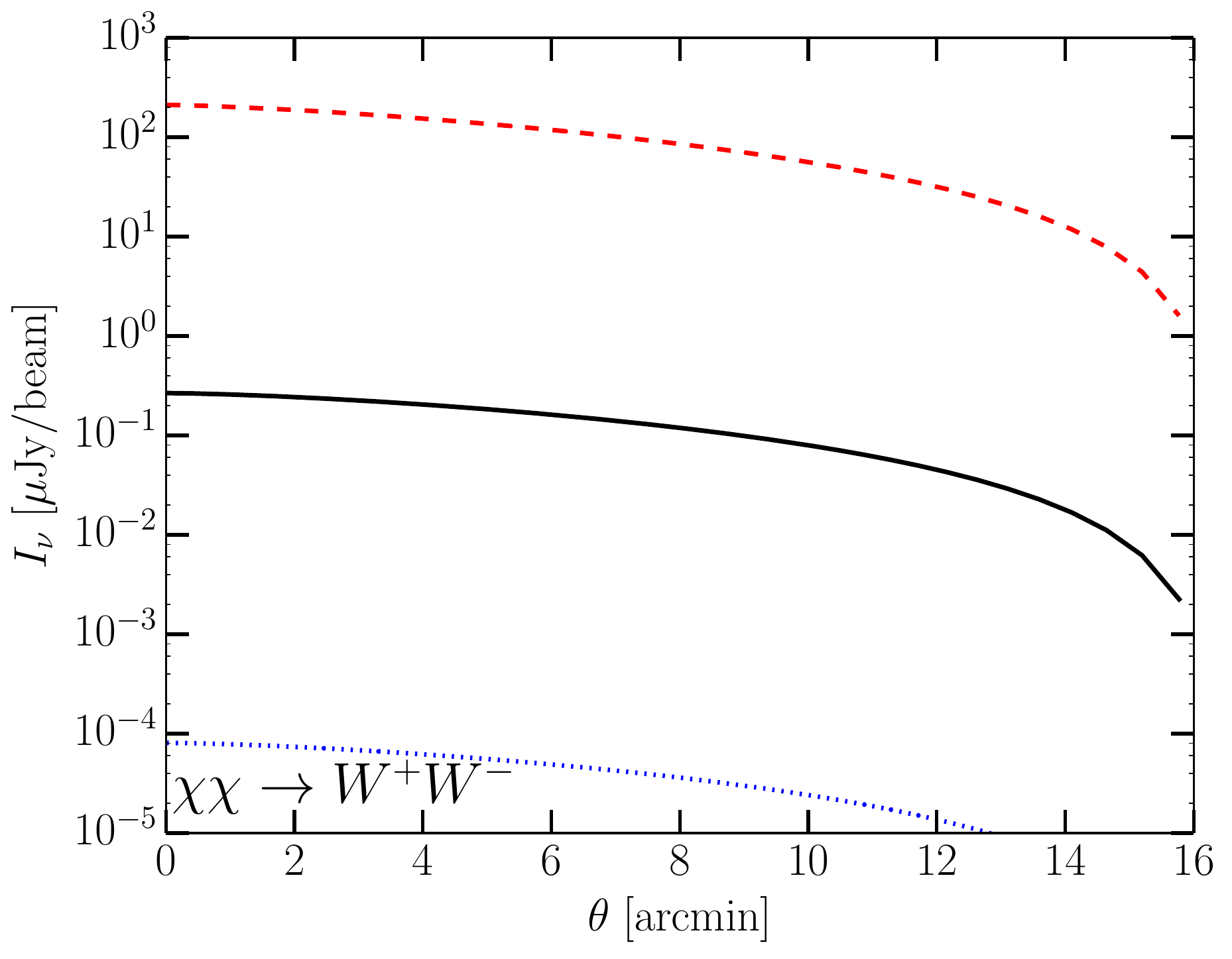}
\includegraphics[width=.49\linewidth]{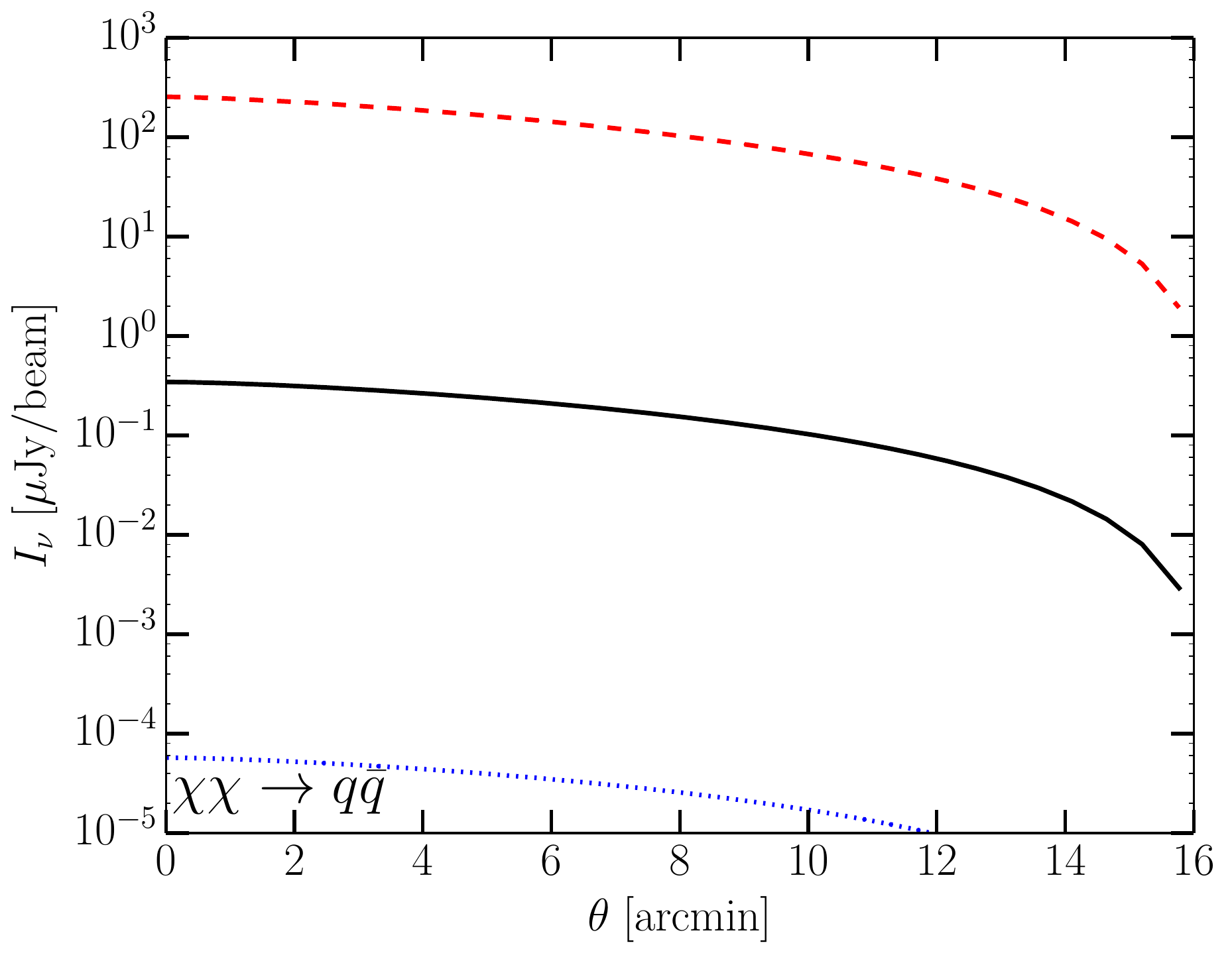}
\includegraphics[width=.49\linewidth]{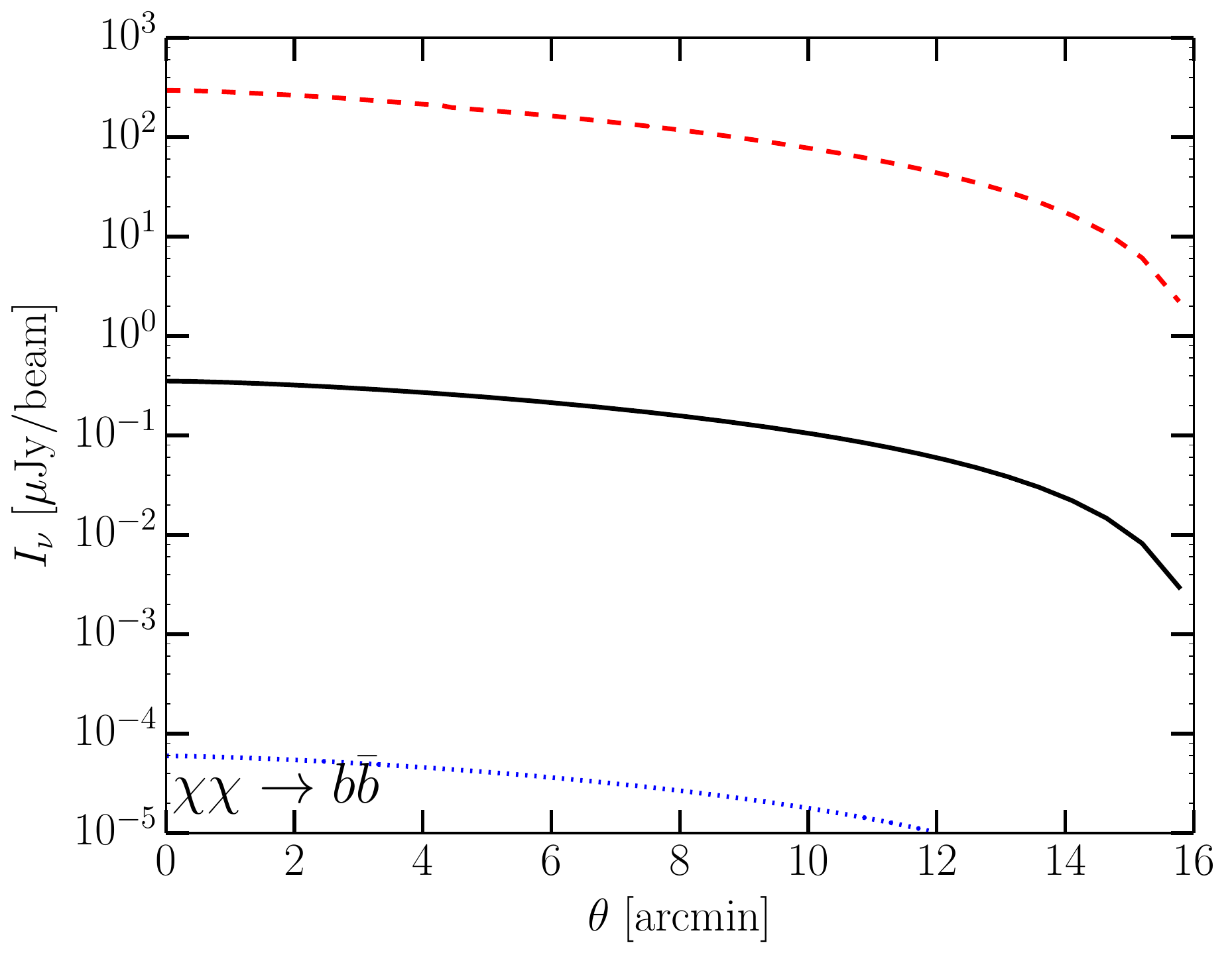}
\caption[...]{Predicted brightness distributions for CVnI for different annihilation channels. Colours and line styles are identical to Fig.~\ref{fig:flux}.}
\label{fig:fluxes}
\end{figure*}

\begin{figure*}
\includegraphics[width=.42\linewidth]{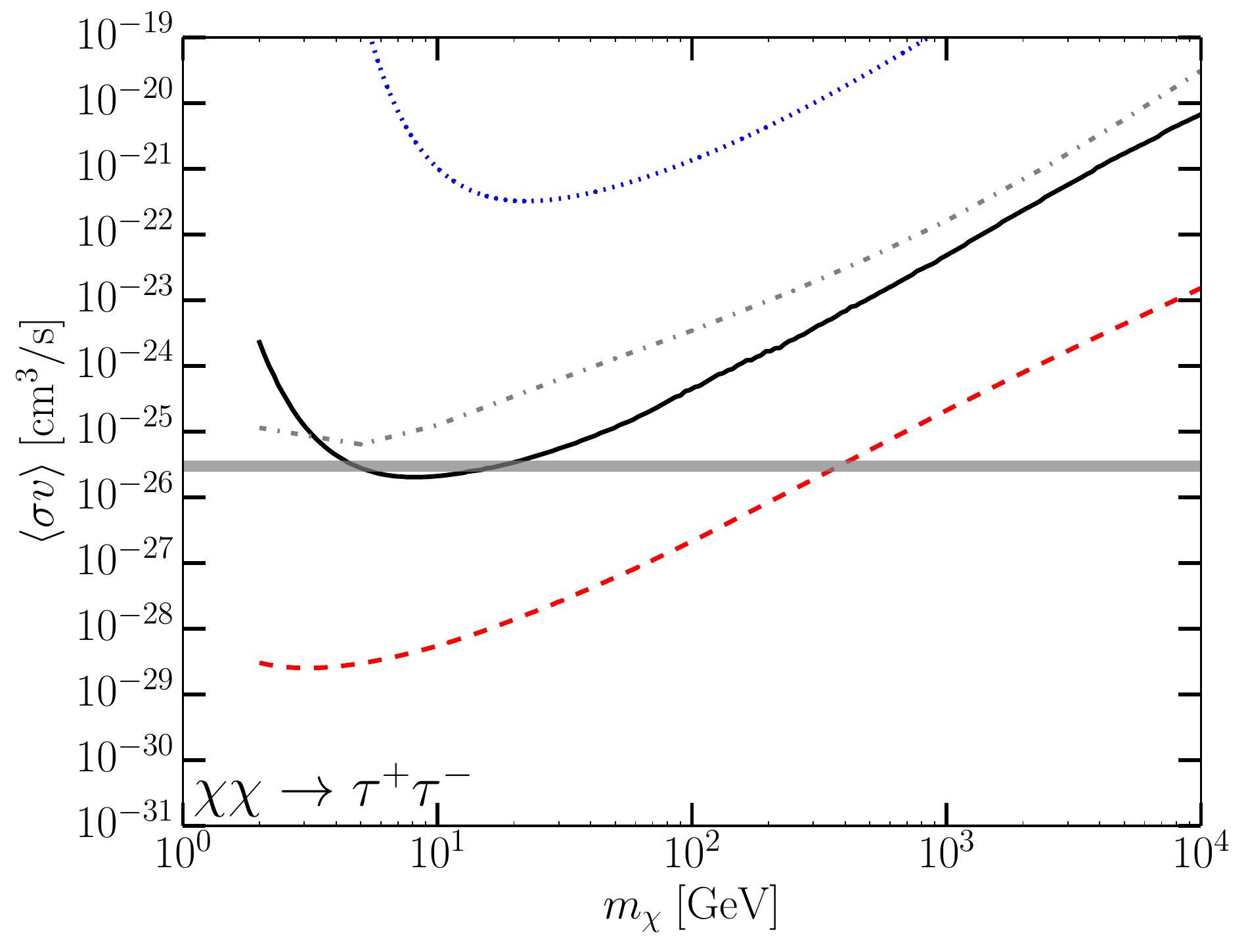}
\includegraphics[width=.42\linewidth]{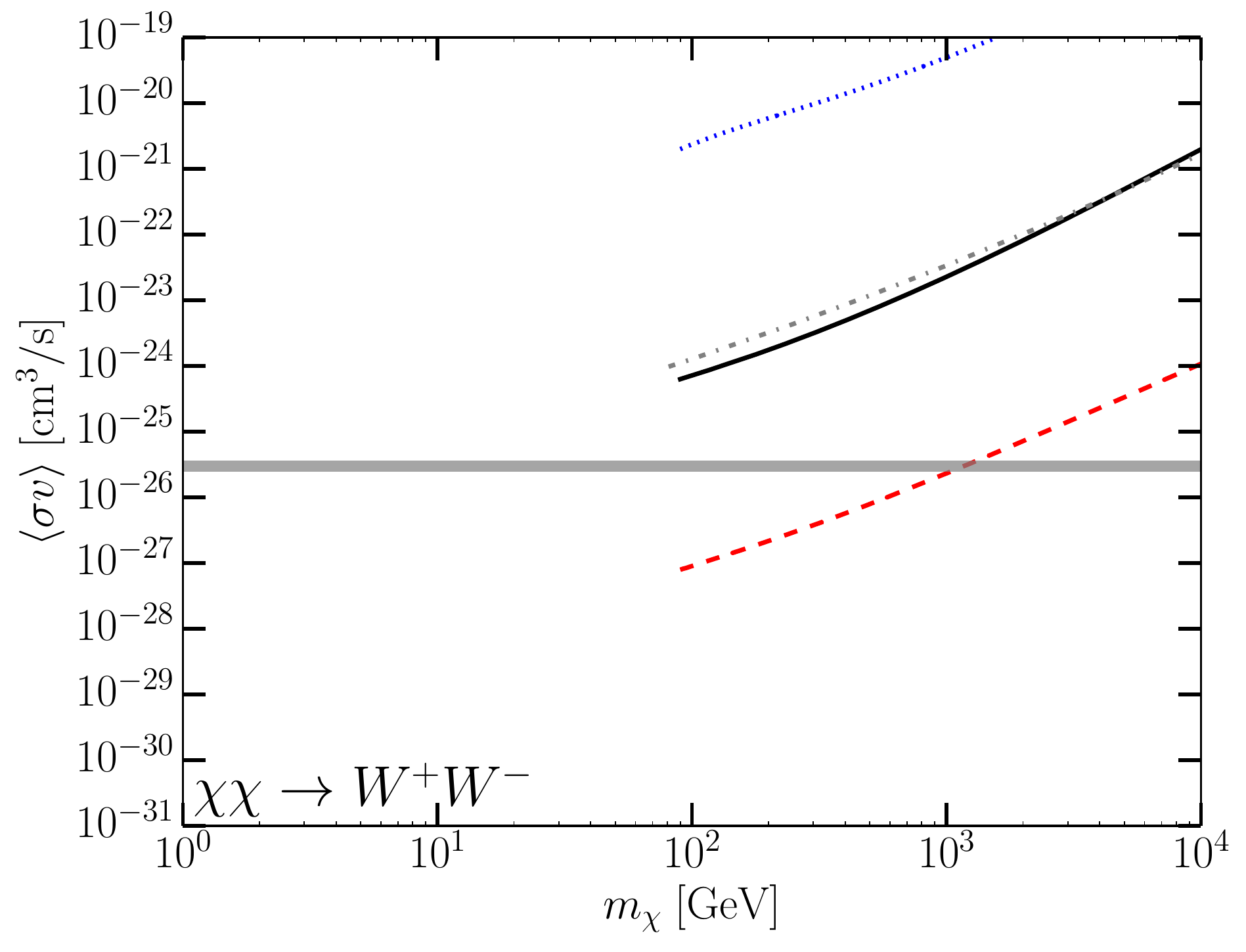}
\includegraphics[width=.42\linewidth]{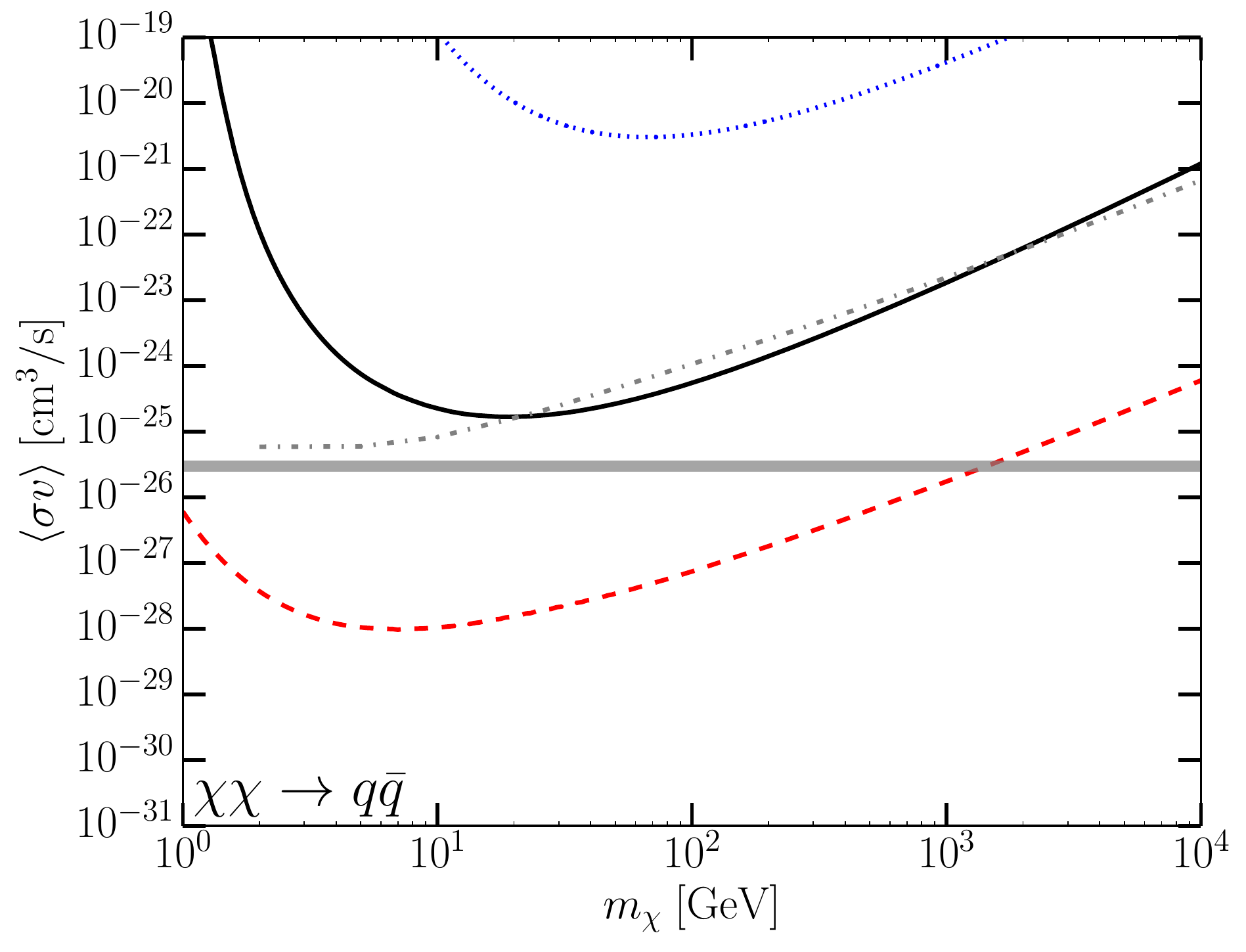}
\includegraphics[width=.42\linewidth]{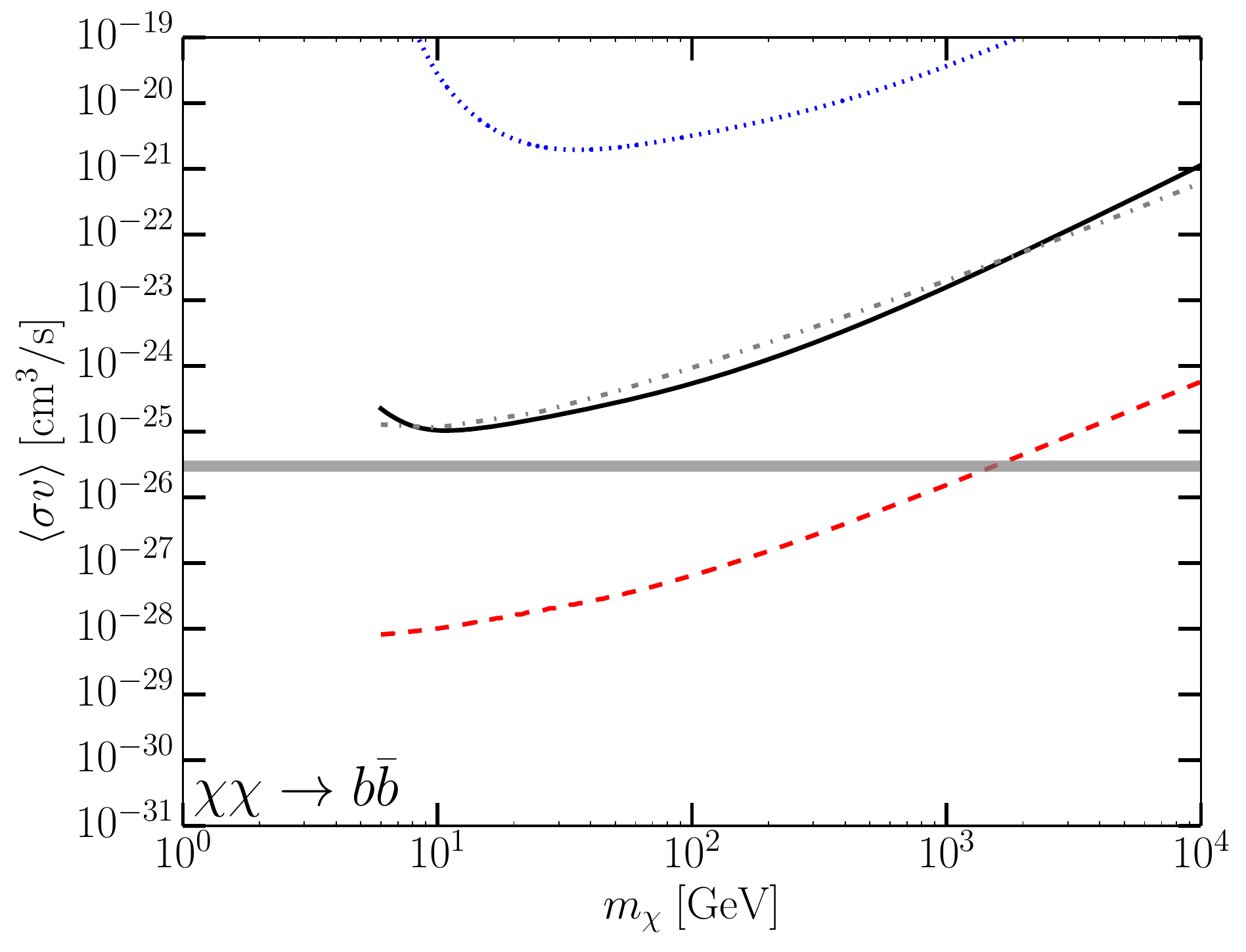}
\caption[...]{LOFAR CVnI limits on the cross section of annihilation of DM into several annihilation channels. Colours and line styles are identical to Fig.~\ref{fig:eelimits}.}
\label{fig:limits}
\end{figure*}

In this Appendix, we include our flux predictions for several annihilation channels (Fig.~\ref{fig:fluxes}) and the limits that we obtain on the DM annihilation cross section into those channels (Fig.~\ref{fig:limits}). Specifically, we considered $\tau^+\tau^-$, $W^+W^-$ $\bar qq$ and $\bar b b$ where $\bar qq$ refers to any neutral pair of light quarks ($q=u,d,s$).
\noindent

\noindent

\section{Further sources of irreducible uncertainties}
\label{sec:appvariations}
\begin{figure}
\center{\includegraphics[width=\linewidth]{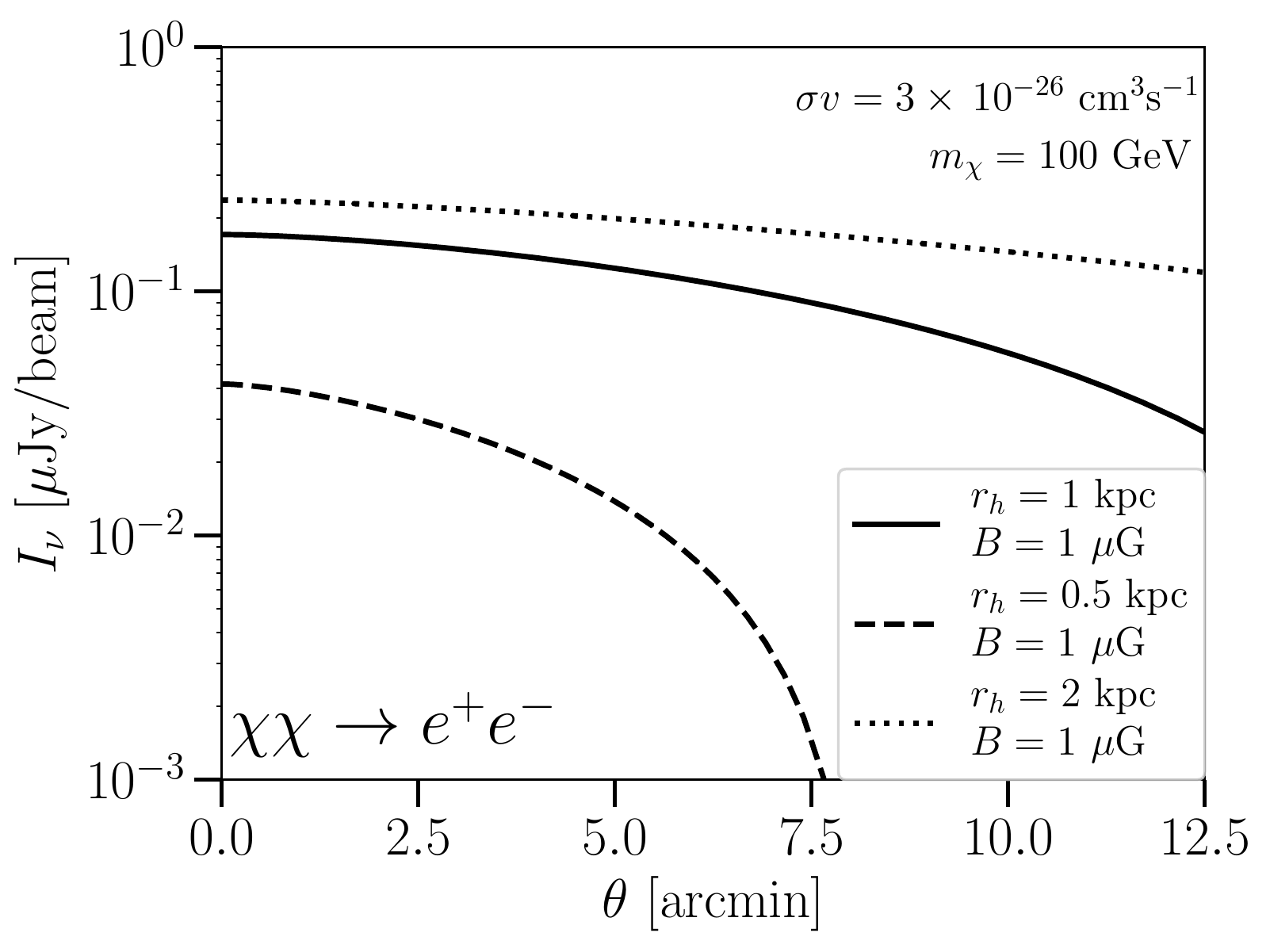}}
\caption[...]{Variations of the parameter $r_h$ in our benchmark model described in the text.}
\label{fig:rh}
\end{figure}

As stressed throughout the main text and visualised in our results (Figs \ref{fig:flux} and \ref{fig:fluxes}), the synchrotron fluxes investigated in this paper are rather uncertain. The chief source of uncertainty is the weakly constrained smooth and turbulent components of the magnetic field in CVnI. In our setup, the uncertainties associated with the smooth component were quantified by {\it directly} varying it in our equations. $B$-field turbulence uncertainties were instead quantified indirectly by altering the normalisation of the diffusion coefficient $D_0$. In addition, one can consider variations of the also turbulence-dependent halo-radius parameter $r_h$. Figure~\ref{fig:rh} shows the results of considering $r_h=r_\star$ and $r_h=4\,r_\star$  instead of the benchmark value adopted in the main text ($r_h=2r_\star$). We observe that, as it was constructed, the extension of the signal scales linearly with $r_h$ and that the $\mathcal O(1)$ variations of $r_h$ we considered yield to $\mathcal O(1)$ effects on our predictions. A similar result is found when even larger values of $r_h$ are considered.

\begin{figure}
\center{\includegraphics[width=\linewidth]{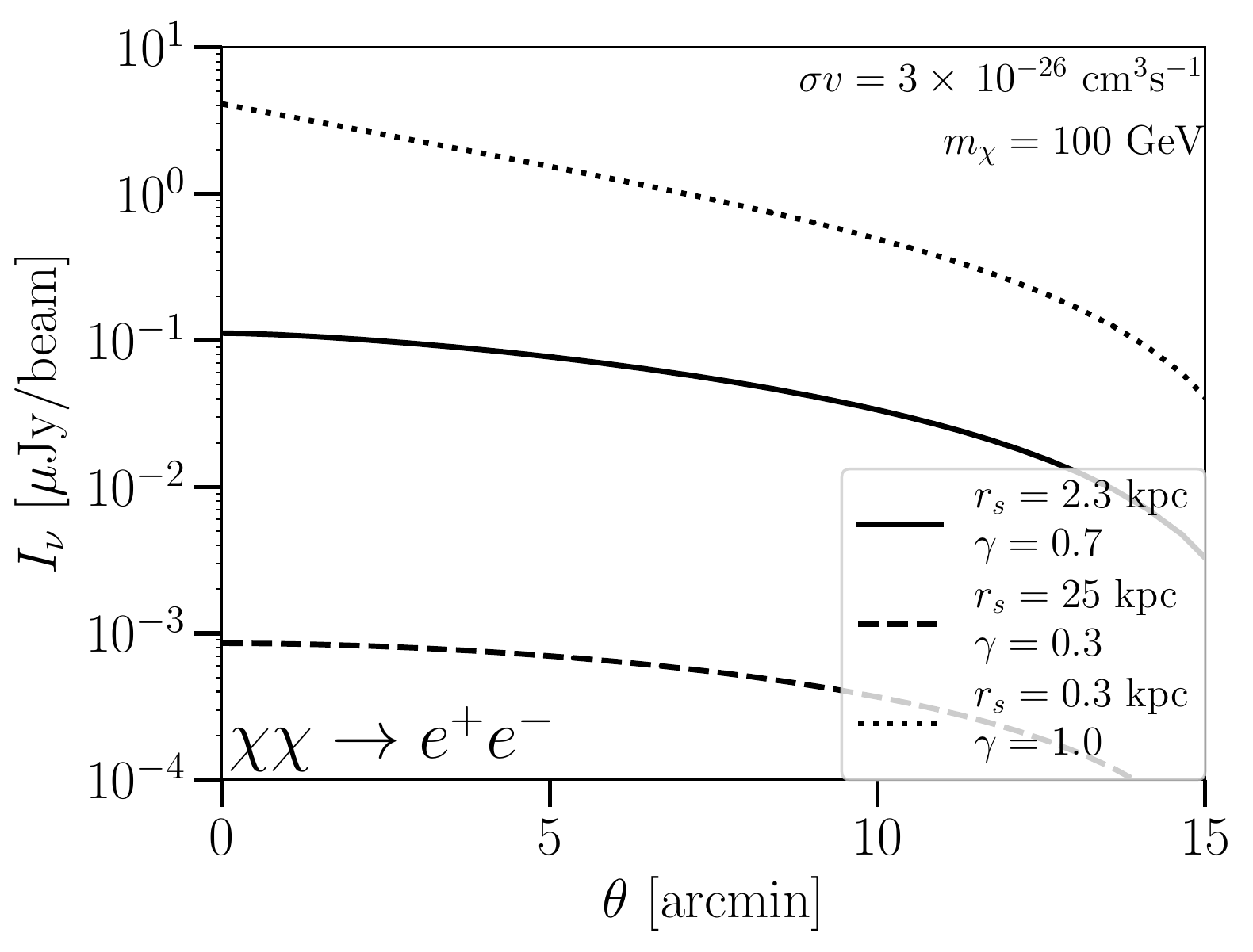}}
\caption[...]{Variations of the parameters $\rho_s$, $r_s$, $\alpha$, $\beta$ and $\gamma$ (only $\gamma$ and $r_s$ shown in legend) in equation~\eqref{eq:density} for the benchmark model in Fig.~\ref{fig:flux}.}
\label{fig:rho}
\end{figure}

Another potentially strong source of uncertainty is the mass function of the dwarf galaxy. Specifically, the parameters entering the DM profile (equation~\ref{eq:density}) are rather unconstrained. Quantifying the impact of these uncertainties on our results is far from straightforward as it would require considering the (six-dimensional) posterior distribution of the kinematic analysis performed in \citet{2015ApJ...801...74G}. We thus estimate the propagated uncertainties on our predictions that are associated with the DM profile, by extrapolating the results of \citet{2015ApJ...801...74G} in their discussion of uncertainties for the prompt gamma-ray spectrum due to DM annihilation ($J$ factors). Specifically, \citet[][their table~2]{2015ApJ...801...74G} quotes 1$\sigma$ uncertanties $\lesssim10^{\,0.37}\approx 2.34$ for CVnI. However, these uncertainties increase for smaller angles \citep[see fig.~7 in][]{2015ApJ...801...74G} probably affecting the $J$ factors (and our extrapolations) by one order of magnitude or more. As an exercise, we consider in Fig.~\ref{fig:rho} two extreme cases where the DM-profile parameters are modified in such a way that, in the benchmark model for the magnetic field's smooth and turbulent components, the synchrotron fluxes are maximised and minimised. These fluxes are separated by several orders of magnitude. Nevertheless, since the statistical significance of the parameter choices for the maximum and minimum fluxes in Fig.~\ref{fig:rho} are unknown to us, the reader should take these uncertainties as order of magnitude estimates only.

\section{Source detection tests}
\label{a:source_detection}

\begin{figure}
\center{\includegraphics[width=0.5\textwidth]{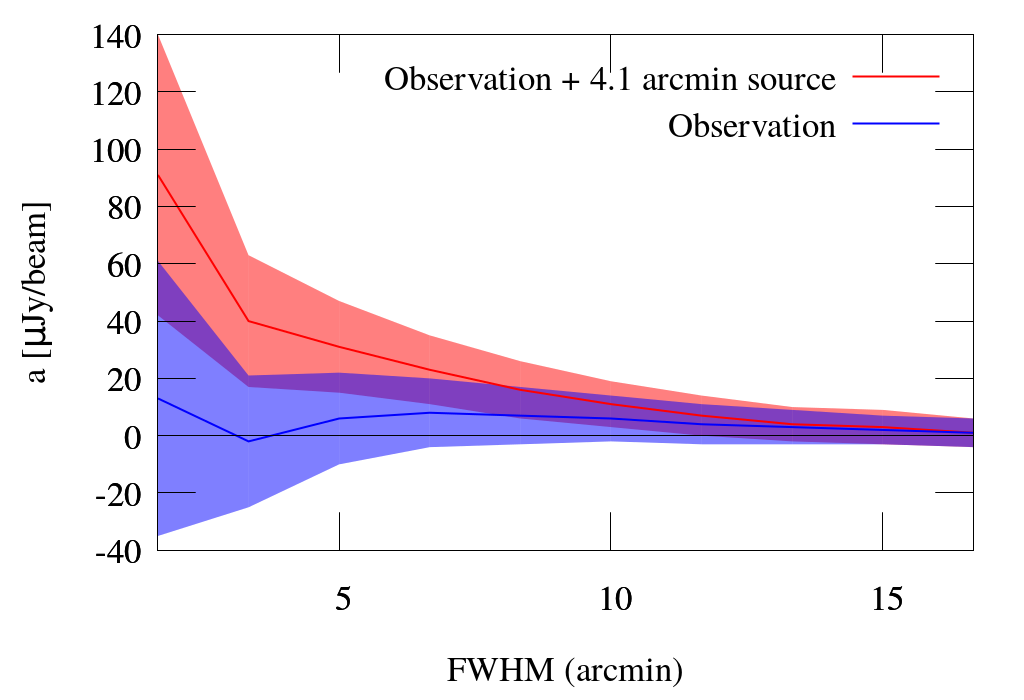}}
\caption[...]{Best-fitting amplitude $a$ for a Gaussian function fitted to the radial intensity profile as function of the assumed value for $b$ (here expressed by FWHM). Lines show the best-fitting amplitudes for the data with a fake 5-mJy source with $\rm FWHM=4.1~arcmin$ (equivalent to an amplitude of $a=32~\rm \mu Jy\, beam^{-1}$) inserted (red) and the control data with no source inserted (blue). Shaded areas indicate 1$\sigma$ uncertainties with 1 degree of freedom.}
\label{fig:sigma_250}
\end{figure}

\begin{figure}
\center{\includegraphics[width=0.5\textwidth]{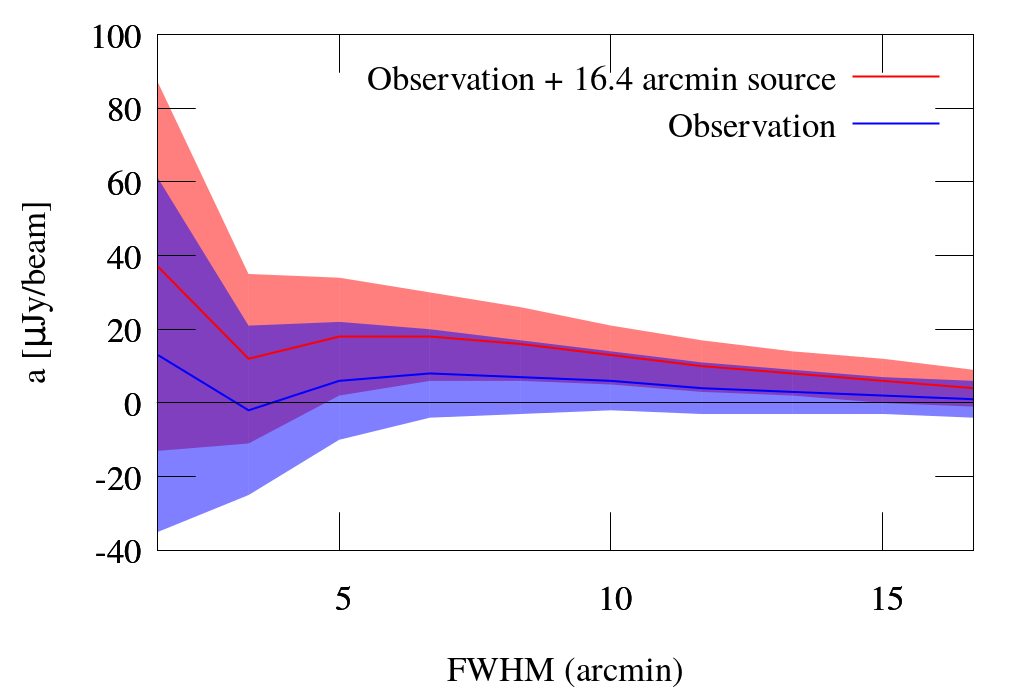}}
\caption[...]{Best-fitting amplitude $a$ for a Gaussian function fitted to the radial intensity profile as function of the assumed value for $b$ (here expressed by FWHM). Lines show the best-fitting amplitudes for the data with a fake 80-mJy source with $\rm FWHM=16.4~arcmin$ (equivalent to an amplitude of $a=32~\rm \mu Jy\, beam^{-1}$) inserted (red) and the control data with no source inserted (blue). Shaded areas indicate 1$\sigma$ uncertainties with 1 degree of freedom.}
\label{fig:sigma_1000}
\end{figure}

% Example table
\begin{table}
	\centering
	\caption{Significance of source detection for inserted fake sources with 4.1 ($\sigma_4$), 8.2 ($\sigma_8$), and 16.4~arcmin ($\sigma_{16}$) FWHM, respectively. For comparison, the data without a fake source ($\sigma_0$) is presented as well.}
	\label{tab:sigma}
	\begin{tabular}{lcccr} % four columns, alignment for each
		\hline
		FWHM & $\sigma_4$ & $\sigma_8$ & $\sigma_{16}$ & $\sigma_0$\\
		(arcmin) & & & &\\
		\hline
		$1.7$   & $1.5$ & $1.5$ &   $0.7$ &	$0.3$ \\
        $3.4$   & $1.6$ & $1.6$ &	$0.5$ &	$0.1$ \\
        $5.0$   & $2.3$ & $2.3$ &	$1.1$ &	$0.4$ \\
        $6.7$   & $2.5$ & $2.5$ &	$1.5$ &	$0.7$ \\
        $8.3$   & $2.3$ & $2.3$ &	$1.6$ &	$0.7$ \\
        $10.0$  & $1.4$ & $2.0$ &	$1.6$ &	$0.8$ \\
        $11.7$  & $1.0$ & $1.6$ &	$1.4$ &	$0.6$ \\
        $13.3$  & $0.7$ & $1.0$ &	$1.3$ &	$0.5$ \\
        $15.0$  & $0.5$ & $0.8$ &	$1.0$ &	$0.4$ \\
        $16.7$  & $0.2$ & $0.6$ &	$0.8$ &	$0.2$ \\
		\hline
	\end{tabular}
\end{table}

In addition to was described in the text of the main paper, we have performed two more tests of inserting fake sources into the $(u,v)$ data. We also inserted a 5-mJy source with $\rm FWHM=4.1~\rm arcmin$ into the data, so that the source can be described by $I_{\nu}=32\exp(-\Theta^2/(2b^2))~\rm \mu Jy\,beam^{-1}$ with $b=1.8~\rm arcmin$, shown in Fig.~\ref{fig:sigma_250}. Similarly, we inserted a 80-mJy source with $\rm FWHM=16.4~\rm arcmin$ into the data, $I_{\nu}=32\exp(-\Theta^2/(2b^2))~\rm \mu Jy\,beam^{-1}$ with $b=7.1~\rm arcmin$, results of which are presented in Fig.~\ref{fig:sigma_1000}. In both cases, the fake source can be detected at $1\sigma$ significance over a range of FWHM values assumed for the fitted Gaussian function. In contrast, the $(u,v)$ data with no source added shows nowhere even a $1\sigma$ detection. Results of these tests are summarised in Table~\ref{tab:sigma}. We find a $2\sigma$ detection for both the $4.1$ and $8.2$~arcmin sources, whereas the $16.4$-arcmin source is detected at $1.6\sigma$ significance. The limit of 2$\times$ the stellar radius for the extent of the DM is an upper limit for us if we adopt the aforementioned $(u,v)$-cut. However, larger $r_h$ also result in higher intensities so that we would be easier able to detect the emission. In contrast, the significance for our data with no source peaks at FWHM=10.0~arcmin with $0.8\sigma$. To summarise, our 2$\sigma$ detection limit
is $32~\mu\rm Jy\, beam^{-1}$ peak intensity with source sizes between $4.1$ and $16.4$~arcmin.

 We also tested the influence of the source detection with {\small PyBDSF} by choosing to include extended sources ({\tt atrous\_do = True}), so that we could get an estimate for the upper limit of the background source contribution. The integrated flux density of CVnI within the $8.5$-arcmin radius is $85.6$~mJy and $6.5$~mJy after standard source subtraction. Hence, the source contribution is $79.4$~mJy for standard source detection; including extended sources, this contribution is $\approx$2~mJy higher. Our manual source detection, where we integrated the flux density in regions around the sources rather than adding up Gaussian components, results in a contribution of $89.5$~mJy. For our fake source detection tests, we have used standard source detection, so that the subtraction of sources is a lower limit.

\bibliography{lofardm}
\bibliographystyle{mnras}

\end{document}